\documentclass[review]{elsarticle}
\pdfoutput=1
\newcommand{\isoverleaf}{1}

\usepackage{tabularx}
\usepackage{lineno}
\usepackage{amsfonts,amssymb,amsmath,exscale}
\usepackage{bm}
\usepackage{subcaption}
\usepackage{xcolor}
\usepackage{import}
\usepackage{gensymb}
\usepackage{upgreek}
\usepackage{verbatim}
\usepackage{xfrac}
\usepackage[locale=UK, unit-mode=text]{siunitx}
\usepackage[utf8]{inputenc}
\usepackage[T1]{fontenc}
\usepackage{nomencl}
\makenomenclature
\usepackage[margin=2.5cm]{geometry}
\modulolinenumbers[5]
\usepackage[colorlinks=true,citecolor=blue]{hyperref}
\usepackage{pgfplots}

\usepackage{adjustbox}
\journal{International Journal of Fatigue}

\ifnum\isoverleaf>0
\biboptions{authoryear}
\else
\biboptions{authoryear}
\fi

\DeclareMathAlphabet{\IKbb}{U}{bbm}{m}{sl}
\DeclareMathAlphabet{\Ikbb}{U}{bbmss}{m}{it}
\DeclareRobustCommand{\KIC}{{\IKbb C}}

\begin{document}

\begin{frontmatter}

\title{
Micromechanical fatigue experiments for validation of microstructure-sensitive fatigue simulation models} 


\author[mymainaddress,myfourthaddress,myfifthaddress]{Ali Riza Durmaz\corref{mycorrespondingauthor}}
\cortext[mycorrespondingauthor]{Corresponding author}
\ead{ali.riza.durmaz@iwm.fraunhofer.de}
\author[mysecondaryaddress,mysixthaddress]{Erik Natkowski}
\author[mysecondaryaddress,mythirdaddress]{Nikolai Arnaudov}
\author[mysecondaryaddress]{Petra Sonnweber-Ribic}
\author[mythirdaddress]{Stefan Weihe}
\author[mysixthaddress]{Sebastian M{\"u}nstermann}
\author[mymainaddress,myfourthaddress]{Chris Eberl}
\author[mymainaddress,myfifthaddress]{Peter Gumbsch}

\address[mymainaddress]{Fraunhofer Institute for Mechanics of Materials, 79108 Freiburg, Germany}
\address[mysecondaryaddress]{Robert Bosch GmbH, Center for Research and Advance Engineering, 71272 Renningen, Germany}
\address[mythirdaddress]{Material Testing Institute, University of Stuttgart, 70569 Stuttgart, Germany
}
\address[myfourthaddress]{Department of Microsystems Engineering, University of Freiburg, 79110 Freiburg, Germany}
\address[myfifthaddress]{Institute for Applied Materials IAM, Karlsruhe Institute of Technology, 76131 Karlsruhe, Germany}

\address[mysixthaddress]{Integrity of Materials and Structures, Steel Institute, RWTH Aachen University, 52072 Aachen, Germany}

\begin{abstract}
Crack initiation governs high cycle fatigue life and is susceptible to microstructural details. While corresponding microstructure-sensitive models are available, their validation is difficult. We propose a validation framework where a fatigue test is mimicked in a sub-modeling simulation by embedding the measured microstructure into the specimen geometry and adopting the experimental boundary conditions. Exemplary, a phenomenological crystal plasticity model was applied to predict deformation in ferritic steel (EN1.4003). Hotspots in commonly used fatigue indicator parameter maps are compared with damage segmented from micrographs. Along with the data, the framework is published for benchmarking future micromechanical fatigue models.

\end{abstract}


\end{frontmatter}


\newcommand{\shearrate}{\dot{\gamma}^{\alpha}}
\newcommand{\shearrateref}{\dot \gamma_0}
\newcommand{\backstress}{\chi_\text{b}^{\alpha}}
\newcommand{\dbackstress}{\dot{\chi_\text{b}}^{\alpha}}
\newcommand{\rss}{\tau^{\alpha}}
\newcommand{\crss}{\tau_\text{c}^{\alpha}}
\newcommand{\nslip}{N_\text{Slip}}
\newcommand{\schmidmat}{{\bm{M}}^{\alpha}}
\newcommand{\schmidmatinter}{\tilde{\bm{M}}^{\alpha}}
\newcommand{\slipnormal}{\bm{n}^{\alpha}}
\newcommand{\slipdirection}{\bm{m}^{\alpha}}
\newcommand{\srs}{m}
\newcommand{\khone}{A_1}
\newcommand{\khtwo}{A_2}
\newcommand{\khthree}{A_3}
\newcommand{\CII}{\KIC^\text{e}_{11}}
\newcommand{\CIZ}{\KIC^\text{e}_{12}}
\newcommand{\CAA}{\KIC^\text{e}_{44}}
\newcommand{\velgrad}{{\bm{L}}^{\text{p}}}
\newcommand{\velgradinter}{\tilde{\bm{L}}^{\text{p}}}
\newcommand{\fipp}{\text{FIP}_\text{p}}
\newcommand{\Dshearmax}{\Delta\gamma_\text{p,max}^\alpha}
\newcommand{\normalstressmax}{\sigma_\text{n,max}^\alpha}
\newcommand{\fipfs}{\text{FIP}_\text{FS}}
\newcommand{\fipw}{\text{FIP}_\text{W}}
\newcommand{\plasticshear}{\gamma_\text{p}^{\slipsysind}}
\newcommand{\tme}{t}
\newcommand{\dtme}{\mathrm{d}t}
\newcommand{\dispthree}{u_3}
\newcommand{\slipsysind}{\alpha}
\newcommand{\kfs}{k}
\newcommand{\surfintervec}{\bm{c}}
\newcommand{\normalsurf}{\bm{n}_\text{surf}}
\newcommand{\slipsysindmax}{\alpha_\text{max}}
\newcommand{\slipnormalmax}{{\bm{n}}^{\slipsysindmax}}
\newcommand{\rotmat}{\bm{R}}
\newcommand{\strain}{\bm{\varepsilon}}
\newcommand{\strainplastic}{\strain_\text{p}}
\newcommand{\dstrainplastic}{\dot{\strain}_\text{p}}
\newcommand{\deq}{d_\text{eq}}
\newcommand{\fipint}{\text{FIP}_\text{int}}
\newcommand{\kint}{k_\text{int}}
\newcommand{\stressgb}{\sigma_\text{n,GB}}
\newcommand{\shearnet}{\gamma_\text{p,net}}
\newcommand{\stressmises}{\sigma_{\text{M}}}

\newcommand{\fippover}{\overline{\fipp}}
\newcommand{\fipfsover}{\overline{\fipfs}}
\newcommand{\fipwover}{\overline{\fipw}}
\newcommand{\fipintover}{\overline{\fipint}}
\newcommand{\stressmisesover}{\overline{\stressmises}}

\section{Introduction} \label{introduction}
A common reason for the failure of mechanical components is the accumulated fatigue damage due to cyclic loading. Considerable efforts are made in the design for reliability. In industry, these efforts predominantly include experimentation and elastic-plastic numerical simulations on the component scale, which leave the material's microstructure aside. However, the microstructure significantly affects early fatigue stages, including crack initiation determining lifetime at small strain amplitudes. More specifically, microstructure-induced life variance and the exact location of crack formation are governed by the reaction of few critical grains. In the past decades, advanced experimental techniques have been developed to observe evolving plasticity under cyclic loading. Thereby, the mechanistic understanding of material degradation is continuously improved and transcribed into materials models. Simulation models help to avoid costly and time-consuming experiments and are therefore of particular interest to the industry.

The crystal plasticity finite element (CPFE) method enables modeling the elastic-plastic deformation behavior of individual grains and clusters in polycrystals. While still limited in simulation domain size, the rapid increase in computational power renders extensive fatigue CPFE calculations feasible, see \cite{boeff2016} and \cite{natkowski_determination_2021}. The transition from predicted mechanical fields to the fatigue state is enabled through metrics for crack initiation, so-called fatigue indicator parameter (FIP) formulations. For various FIP definitions, see e.g. \cite{schaefer2019b}.

\cite{manonukul2004} propose the plastic slip accumulated over each slip system as a crack initiation criterion. Based on the work of \cite{fatemi1988}, \cite{bennett2003} propose a FIP, where the irreversibility of the plastic strain and the normal stress applied on the critical slip plane is assumed to be a driving force for crack initiation. 
For a discussion of different FIP formulations, it is referred to \cite{hochhalter2010}. While these approaches can predict damage location and initiation life and are supported by a growing research community, they demand microstructural characterization and experimental benchmarks. However, an established and standardized validation procedure for micromechanical fatigue simulations is not yet available.

The validation of CPFE approaches by experimental means often involves high-resolution digital image correlation (HR-DIC) in conjunction with mechanical testing. Several of these approaches were reviewed in \cite{Pokharel2014}. \cite{guan2017} investigated single and oligo crystal nickel material with a three-point beam bending under conditions of cyclic loading. They identify slip activation with SEM images and quantify developing strain fields and localization during fatigue with HR-DIC. CPFE models are applied on microstructures under consideration of the fatigue loading such that grain-by-grain comparisons of slip can be carried out. Similarly, HR-DIC at micro tensile test specimen of a polycrystalline Ni-base superalloy has been performed by \cite{eastman2018} to benchmark a CPFE model of this material. In \cite{Zhang2018} slip bands observed in HR-DIC after tensile tests yielded good agreement with CPFE simulations concerning slip trace orientations. In prior work, \cite{Zhang2016} quantitatively compared CPFE strains in a nickel alloy grain cluster with HR-EBSD after fatigue testing. A coupling between in-situ micropillar compression of a lamellar TiAl alloy and CPFE modeling is described by \cite{chen2019}.

The aforementioned approaches have in common that few grains are analyzed, implying a small statistical basis. Furthermore, fatigue experiments are mainly performed in the low cycle fatigue (LCF) regime, while the very high cycle (VHCF) regime arguably is less understood and governed by a more pronounced impact of microstructure. The literature benchmarks fatigue phenomena predominantly up to slip band emergence utilizing experimental, image correlation-derived strain fields. While representing essential validations, slip bands do not necessarily culminate in crack initiation followed by short crack propagation.

To complement these approaches, bending resonant fatigue experiments up to short crack growth are accompanied by spatial distortion-corrected EBSD scans covering the entire highly loaded mesoscale specimen surface. The correction of such an EBSD scan facilitates its accurate alignment with damage data across the whole region of interest. Therefore, the number of considered grains can be substantially increased, and the statistical basis thereby significantly improved. While similar multimodal frameworks for large specimen areas exist \cite{charpagne2020automated, chen2018high}, they, to the best of our knowledge, have not been applied to statistically validate the damage localization of computational models. Instead of attempting validation of strain fields through DIC, SEM images were semantically segmented with the help of an accordingly trained convolutional neural network (CNN) to find locations of surface plasticity and cracks, see \cite{Thomas2020}.

The resonant bending fatigue setup by \cite{Straub2015} relies on planar mesoscale specimens with optimized geometry for a large uniform stress state at the specimen surface. The highly loaded zone typically covers an area of $500\times1000\,\upmu$m$^2$ which still can be covered by CPFE simulations for the microstructure at hand. Corresponding simulations depend on well-aligned geometry, boundary conditions, and microstructure data. Furthermore, the validation requires reference damage localization maps and cyclic damage evolution information. Thus, the data registration procedure presented in \cite{Durmaz2021}, where multiple heterogeneous data sets were spatially aligned, is employed here. Namely, in-situ light optical image series, high resolution stitched secondary electron (SE2) images, and deduced damage maps were aligned with the microtexture from 2D EBSD. The registered data is used for setting up a sub-modeling approach, where the macro model comprises the whole bending beam geometry, and the available EBSD data, confined to the highly loaded region, defines the CPFE submodel within the beam geometry. The experimental boundary conditions are translated onto the microstructure through the macro model. This unidirectional coupling allows the computation of the micromechanical stress and strain response as well as deduced FIP metrics for the actual microstructure. Such FIP maps and specifically contained hotspots are compared to damage locations identified by the aforementioned CNN for semantic segmentation. Thereby, the FIP's suitability for predicting locations of surface plasticity traces and cracks on the microstructure scale is assessed.

The experimental data for model validation, micromechanical simulation data, and the Abaqus sub-modeling routine for easy integration of user subroutines are accessible through our research data repository \cite{fordatis_repo}  and a GitHub repository \cite{github_repo}.
The latter includes detailed instructions on how to use the data and set up simulation models based upon them.
It is hoped that other researchers will find this helpful to test their micromechanical fatigue models' validity and that it will lay the foundation for a standard benchmarking approach of such.


\section{Material and experimental} \label{experimental}

The experimental procedure applied is outlined in detail in \cite{Durmaz2021}. An abridged version is presented below.

\subsection{Specimen preparation and complementing microscopy}
\label{sec:Specimen_microscopy}
The investigated material is a ferritic stainless steel EN 1.4003 (AISI 3Cr12). Table \ref{chemical_comp} shows the nominal values of the alloying elements. The as-received rod material underwent hot rolling, grinding, cold drawing, and annealing. Disks of 0.7\,mm thickness were extracted from the rod by electrical discharge machining, from which, in turn, mesoscale specimens were laser cut. Finally, state-of-the-art grinding, Struers A2 electropolishing (electrolyte composed of perchloric acid, 2-Butoxyethanol, ethanol, and water), and diamond particle/OPS polishing accomplished the specimen surface finish.
 
\begin{table}[hbtp]
	\centering
	\begin{tabular}{lccccccccc}
		\hline
		Material & C & Si & Mn & P & S & N & Cr & Ni & Mo \\ \hline
		\hline \vspace{-4ex} \\ 
		\textbf{1.4003} $\;$ & 0.013 & 0.67 & 1.08 & 0.018 & 0.021 & 0.013 & 11.9 & 0.43 & 0.33 \\
		\hline
	\end{tabular}
	\caption{Chemical composition of ferritic stainless steel 1.4003 in weight-\%.}
	\label{chemical_comp}
\end{table}

\begin{figure}[htbp]
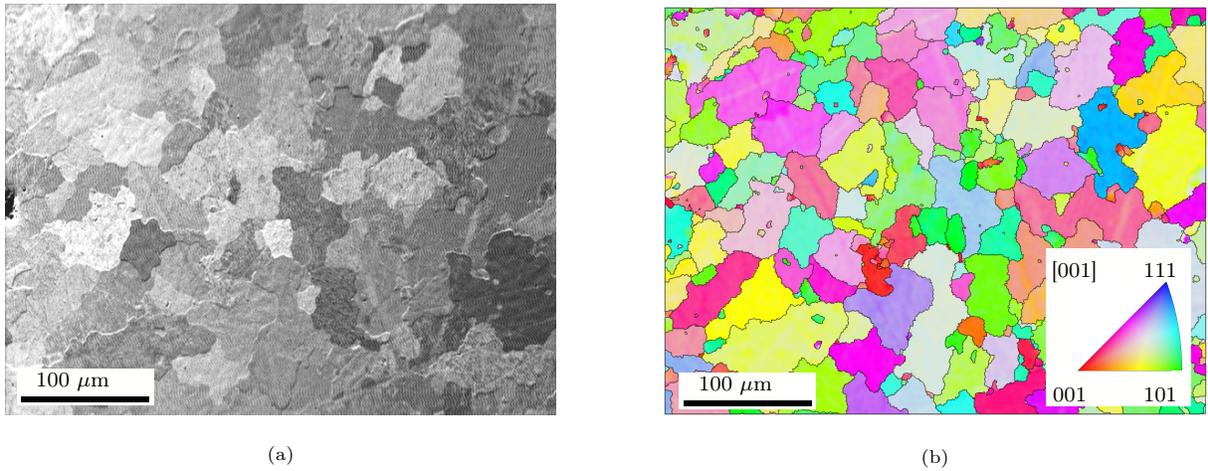

	\centering
	\begin{subfigure}{0.45\textwidth}
		\def\svgwidth{1.0\textwidth}
		\footnotesize
		\input{figures/microstructure/rem_render.pdf_tex}
		\caption{}
	\end{subfigure}
	\hspace{10mm}
	\begin{subfigure}{0.45\textwidth}
		\def\svgwidth{1.0\textwidth}
		\footnotesize
		\input{figures/microstructure/ebds_render.pdf_tex}
		\caption{}
	\end{subfigure}
	\caption{Microstructure analysis of the 1.4003 ferritic steel microstructure. A SEM image is shown in (a) and an [001] EBSD map in inverse pole figure color coding illustrated in (b).}
	\label{fig:microstruture}
\end{figure}

Supplementary measurements included topography-sensitive SEM secondary electron (SE2) imaging before (1) and after (2) fatigue, as well as an EBSD scan before fatigue (3). Each of these techniques was applied on the highly loaded regions and both specimen sides. A cut-out of (1) and (3) is shown in Figure \ref{fig:microstruture} a) and b), respectively. These measurements utilized a Zeiss Supra 40VP equipped with an EDAX TSL EBSD system. For EBSD data, the working distance (WD), scan step size, and aperture were chosen to be 18\,mm, 0.6\,$\upmu$m, and 60\,mm, respectively. The EBSD image reveals rather large grains with an average equivalent diameter of $\deq \approx 25.5$\,$\upmu$m and single grains up to 100\,$\upmu$m in size. The specimen alignment relative to the rod results in slightly elongated grains in the out-of-plane direction of the specimen (not shown), where the grain size reaches up to 130\,$\upmu$m. There are manganese sulfide inclusion lamellae (type II and type III) \cite{farrar1974inclusions} which reach length of up to 50\,$\upmu$m and are oriented along the the specimen out-of-plane direction.

The acquisition of the three data sets was motivated by the need for an undistorted and undeformed reference indicating prior defects (1) as well as high-resolution damage (2) and microtexture (3) information. Former allows for a spatial distortion correction of EBSD data proposed by \cite{Nolze2007, Kapur2000, Wu2002}. To reduce imaging-induced distortions in the SEM reference (1), high magnifications, long dwell times, and low working distances were utilized as suggested in \cite{Kamm2013}. For the same reason and to ensure appropriate imaging of pores and micromechanical damage, both SE2 scans were performed as stitching scans. The detailed correction approach is delineated in \cite{Durmaz2021}.

One difficulty in mapping EBSD onto other image-based modalities is the contained spatial distortions due to the 70\degree{} specimen tilt angle in EBSD measurement \cite{Nolze2007, Nolze2006}. In our case, further distortions are superimposed that arise due to the fabrication-induced specimen surface curvature. In addition to the spatial distortions, the orientation measurements are affected if large specimen surfaces are scanned, see \cite{Ram2015, Nolze2007}. Primarily, the spatial distortions in the EBSD data impede the correct assignment of damage derived from rather undistorted SEM images to its correct grain in the distorted microstructure from EBSD. This is intensified by the fact that extrusions and cracks frequently emerge at grain boundaries, see \cite{Sangid2013, Polak2003}. Therefore, the correction of large EBSD scans represents a fundamental requirement to perform reliable microstructure-property relationship analysis.

\subsection{Mesomechanical fatigue testing}
\begin{figure}[htbp]
	\centering
	\footnotesize
	\import{figures/experimental/}{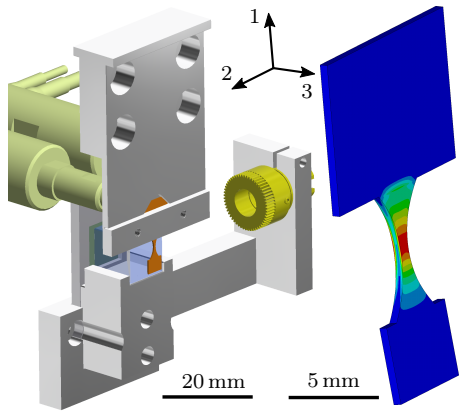} 
	\caption{Overview of the fatigue setup illustrating the arrangement of piezo acuators (khaki), micro sample (orange), laser (yellow) and the position sensitive device (green). The right part shows the specimen in a loaded state and the corresponding von Mises stress state.}
	\label{fig:Overview_setup}
\end{figure}

The experimentation was performed with a fatigue setup that enables resonant multiaxial loading based on \cite{Straub2015} and described in \cite{Durmaz2021}. In this work, the applied loading type is bending. The setup uses mesoscale specimens with a small, highly loaded volume in conjunction with a sophisticated control mechanism. Therewith, high sensitivity regarding initial fatigue states such as extrusion formation and crack initiation is achieved. This is enabled by a closed-loop control that adjusts the excitation frequency to adapt for degradation-induced specimen resonant frequency changes. Further, the setup is equipped with a stroboscope and camera system to capture local damage evolution in an in-situ light optical image series. Figure \ref{fig:Overview_setup} depicts the fatigue setup without the additional optics and a specimen-scale bending-induced von Mises stress distribution. 

A fully reversed loading (load ratio of $R$ = -1) is applied, where the rotation angle at the unconstrained beam end acts as the controlled parameter determining the load. The specimen geometry results in a resonant frequency of $f_0$ $\approx$ 2000\,Hz. Attaining $10^9$ cycles or a relative frequency decrease of $\Delta f/f_0 = 0.1\%$ pose stopping criteria for the test. Before the fatigue experiment, the control parameter equivalent to a von Mises stress amplitude of 270\,MPa is computed by linear elastic finite element simulations.

\subsection{Multimodal image registration} \label{image}
Spatial data alignment is a prerequisite to exploit the synergies of the heterogeneous data acquired from the different imaging techniques. This process is denoted as multimodal image registration and involves challenges such as different fields of views, resolutions, contrast (e.g., topography and orientation contrast), and more. For this work, the objective is to perform and validate CPFE sub-modeling simulations.
Therefore, it is necessary to embed the EBSD-derived partially available microstructure information in the idealized specimen geometry and link them to the resulting cyclic damage evolution data and high-resolution damage localization maps. For this purpose, data from the multiple sources (s) were partially transferred to an intermediate target (it) before mapping to the target (t). Specifically, the idealized geometry (s), in-situ light optical image series (s), stitched SE2 before (it) and after fatigue (s), and a damage segmentation mask derived from the latter (s) were aligned with the EBSD scan (t). In this multi-stage registration procedure, all source data sets were first transformed to the intermediate target, the comparatively undistorted and undeformed stitched SE2 SEM image before fatigue. Subsequently, a transformation aligns all data sets with the target modality. Since minimal non-linear spatial distortions were achieved by appropriate imaging conditions and a single specimen holder throughout most of the fatigue and analytical process chain, linear affine (or alternatively rigid/similarity) transformations were adequate to achieve this alignment. The necessary transformation matrices were computed by feature-based, and intensity-based registration techniques, see \cite{Durmaz2021}. On the other hand, the latter transformation to the EBSD target modality demands distortion correction due to its inherent spatial distortions described in Section \ref{sec:Specimen_microscopy}. Hence, this alignment is achieved through an approximate affine transformation followed by an elastic transformation correction of EBSD data using point correspondences and b-spline regularization, see \cite{Sorzano2005}. In all transformation instances, owing to the feature sparsity of the polished surfaces, automatic feature detection and matching of point correspondences proved to be complicated. Therefore, features related to pores and particles common to various modalities enabled the manual feature selection and matching and thus computation of the transformations. Ultimately, the heterogeneous data were spatially aligned, and imaging-based distortions were removed to a large extent.

\section{Micromechanical fatigue simulation} \label{modeling}
The model framework consists of a CPFE model to consider the material's microstructure through grain orientations. Fatigue indicator parameters (FIPs) are introduced to quantify material fatigue during cyclic loading. These parameters are calculated from micromechanical stresses and strains obtained during the CPFE simulation. To validate the CPFE model, a sub-modeling approach is incorporated to capture realistic boundary conditions matching the experiments.

\subsection{Crystal plasticity finite element model}
A phenomenological and isothermal CPFE approach considering small strains is chosen to model elastic-plastic deformation behavior under cyclic loading for polycrystalline materials. As done by \cite{roters2010} length scale dependencies are omitted in this local formulation. A brief overview of the model is given in the following, while details can be found in \cite{kuhn_identifying_2021} and \cite{wicht_efficient_2020}. 

Following \cite{asaro1977}, plastic deformation is assumed to be only caused by dislocation slip. For the BCC lattice the twelve $\{110\}$$\langle 111 \rangle$ slip systems are taken into account, which are preferentially activated, see \cite{courtney2005}. The plastic velocity gradient is thus defined as superposition of all shear rates $\shearrate$ on all corresponding slip systems $\slipsysind$ as
\begin{equation} \label{eq:velocity_gradient}
	\dstrainplastic = \sum^{\nslip}_{\slipsysind=1} \shearrate \schmidmat \, ,
\end{equation}
with the Schmid matrix $\schmidmat = \slipdirection \otimes \slipnormal$ as a dyadic product of the projection on plane $\slipnormal$ and into slip direction $ \slipdirection$. 

The evolution of the shear rate is adapted from \cite{rice1971} and \cite{hutchinson1976} and applied in a modified form of \cite{cailletaud1992} including kinematic hardening. It reads
\begin{equation} \label{eq:slip_rate}
	\shearrate = \shearrateref
	\Big \lvert \frac{\rss - \backstress}{\crss} \Big \rvert^\srs \, \text{sign}(\rss - \backstress) \, ,
\end{equation}
where $\crss$ is the critical resolved shear stress, 
$\shearrateref$ denotes the reference shear rate, and $m$ the shear rate sensitivity exponent, which are material properties. The resolved backstress on each glide system to account for kinematic hardening is described by $\backstress$.

The incorporation of kinematic hardening is necessary for the modeling of fatigue properties during cyclic loading. A kinematic hardening model for the evolution of the resolved backstress proposed by \cite{ohno_kinematic_1993} is applied and reads 
\begin{equation}
	\dbackstress = \khone \shearrate - \khtwo \left( \frac{\lvert \backstress \rvert}{\sfrac{\khone}{\khtwo}}  \right)^{\khthree} \backstress \lvert \shearrate \rvert \, ,
\end{equation}
with the material dependent parameters $\khone$, $\khtwo$, and $\khthree$. Implemented in a user-defined material subroutine (UMAT) of ABAQUS for user defined material models, the above introduced CPFE model provides a tool to describe micromechanical deformation under cyclic loading. The set of material parameters for the investigated ferritic steel 1.4003 is adopted from \cite{natkowski_fatigue_2021} and listed in Table \ref{cpfem_parameter}. 
\begin{table}[htbp]
	\centering
	\begin{tabular}{|c|l|l|c|}	
	\hline
	Parameter & Name & Value & Unit \\ \hline
	\hline \vspace{-3.5ex} \\ 
	$\CII$ & Elastic constant                   & 263.7 & GPa \\
	$\CIZ$ & Elastic constant                   & 120.5 & GPa \\
	$\CAA$ & Elastic constant                   & 79.1 & GPa \\	
	$\nslip$ & Number of slip systems           & 12 & - \\	
	$\shearrateref$ & Reference shear rate      & 0.001 & 1/s \\
	$\srs$ & Strain rate sensitivity exponent   & 100 & - \\
	$\crss$ & Critical resolved shear stress    & 119.8 & MPa \\			
	$\khone$ & Kinematic hardening 1 $\qquad$   & 25550 & MPa \\
	$\khtwo$ & Kinematic hardening 2 $\qquad$   & 1353 & - \\			
	$\khthree$ & Kinematic hardening 3 $\qquad$   & 8 & - \\			
	\hline
\end{tabular}
	\caption{Material parameters for the CPFE model, according to \cite{natkowski_fatigue_2021}.}
	\label{cpfem_parameter}
\end{table}
%
\subsection{Fatigue indicator parameters}
Following \cite{przybyla2010}, irreversible plastic slip on the microscale causes the initiation of fatigue damage. A framework for the numerical tracking and prediction of fatigue damage in nickel-based alloys was proposed in \cite{shenoy2007, przybyla2010, castelluccio2014meso}. As a micromechanical metric for the local degradation state and a driving force for fatigue crack initiation, FIPs were introduced. FIPs are derived from stress and strain fields on the microstructure during cyclic loading. Multiple FIPs are summarized in \cite{hochhalter2010}. 

\cite{manonukul2004} and \cite{mcdowell2010} propose the accumulated plastic strain as FIP. The slip rate introduced in (\ref{eq:slip_rate}) quantifies the activity of plastic gliding on each slip system $\slipsysind$.
The accumulated plastic slip is calculated from the resulting overall plastic strain rate. It describes the amount of accumulated plastic deformation a material has experienced during the cyclic loading history at a local material point. 
It reads
\begin{equation} \label{eq:acc_pl_strain} 
	\fipp  = \int_{\text{cycle}} \sqrt{ \tfrac{2}{3} \, \dstrainplastic : \dstrainplastic} \, \dtme \, .
\end{equation}

A second approach proposed by \cite{fatemi1988} considers the normal stress on the plane of maximum shear strain range. The FIP combines the maximum shear strain range $\Dshearmax$ and the maximum normal stress on the plane of maximum shear strain range $\normalstressmax$. This critical plane approach is determined as
\begin{equation} \label{eq:fatemie_socie} 
\fipfs = \max_{\alpha\in\left(1,...,12\right)} \left[\frac{\Dshearmax}{2} \Big(1+ \kfs \frac{\normalstressmax}{\crss} \Big) \right]
\end{equation}
with
\newcommand{\tmemod}{\tilde{\tme}}
\begin{equation} \label{eq:fatemie_socie2} 
\Dshearmax = \max_{\tmemod}{\Big( \int_{\text{cycle start}}^{\tmemod} \dot \plasticshear ~ \dtme \Big)} - \min_{\tmemod}{\Big( \int_{\text{cycle start}}^{\tmemod} \dot \plasticshear ~ \dtme \Big)} \, ,
\end{equation}
where $\crss$ is the critical resolved shear stress and $\kfs$ is a constant set to 1.0 controlling the influence of the normal stress. For a discussion of this definition and the value of $\kfs$, see \cite{boff_micromechanical_2016} and \cite{castelluccio_study_2012}.

A FIP based on a dissipation energy was introduced by \cite{korsunsky2007}. Here, the amount of energy dissipated over all slip systems under cyclic loading is considered as a measure for fatigue and the FIP is defined as
\begin{equation} \label{eq:energy_criteria} 
\fipw = \int_{\text{cycle}} \, \sum^{\nslip}_{\slipsysind=1} \rss ~ \shearrate \, \dtme \, .
\end{equation}
The implementation in the UMAT framework is similar to the determination of the accumulated plastic slip.
\cite{boeff2017} and \cite{schaefer2019b} compare different FIPs and their suitability for numerical prediction of fatigue crack initiation life.

Since the aforementioned FIP are primarily applicable for transgranular crack initiation but also intergranular cracks are observed, a intergranular FIP
\begin{equation}
\fipint = \shearnet \left(1+\kint\frac{\stressgb}{\crss}\right)
\end{equation} 
is employed. In contrast to the intergranular FIP proposed by \cite{przybyla_microstructure-sensitive_2013}, the critical resolved shear stress is utilized instead of the yield stress, as proposed by \cite{boff_micromechanical_2016}.
$\shearnet$ is the maximum net plastic shear strain among all slip planes, $\stressgb$ is the average peak normal stress on the grain boundary and $\kint$ is a parameter, which is chosen to be $\kint = 1.0$ in analogy to the Fatemi-Socie FIP. 

While aforementioned hardening and FIP formulations were shown to achieve decent damage localization capabilities in LCF settings, their applicability to VHCF is largely unknown. 

%
\subsection{Sub-modeling approach}
\begin{figure}[htbp]
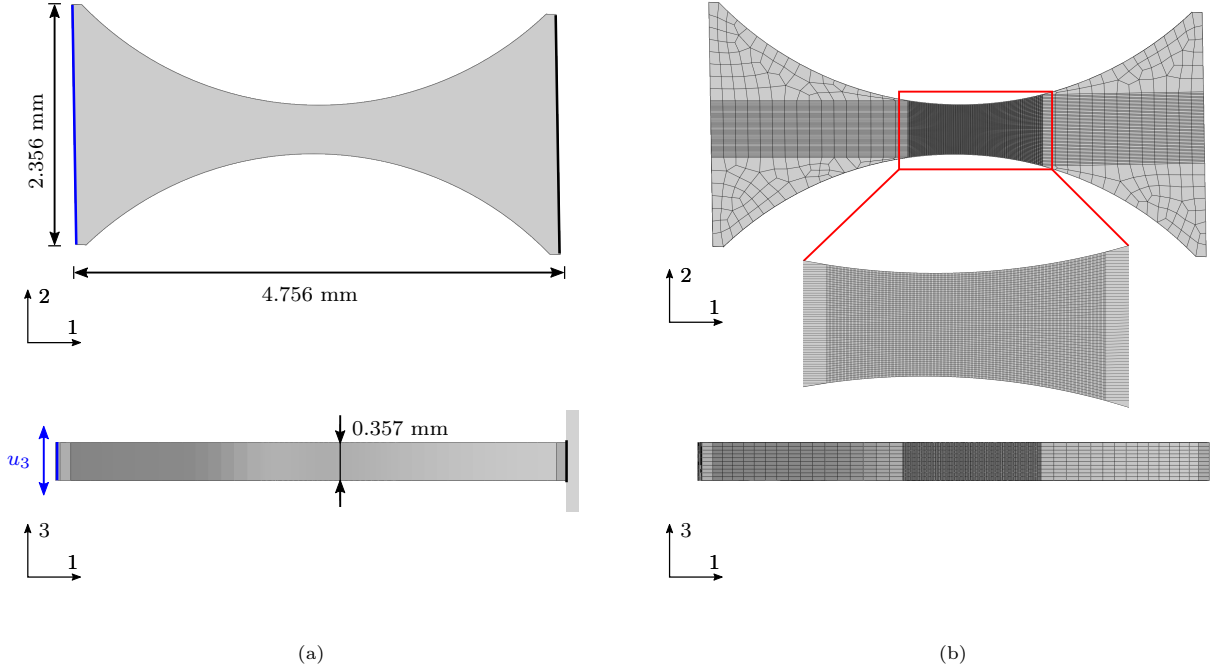

	\centering
	\begin{subfigure}{0.47\textwidth}
		\def\svgwidth{1.0\textwidth}
		\footnotesize
		\import{figures/macro_geom/}{macro_geom_render.tex}
		\caption{}
	\end{subfigure}
	\hspace{5mm}
	\begin{subfigure}{0.47\textwidth}
		\def\svgwidth{1.0\textwidth}
		\footnotesize
		\import{figures/macro_geom/}{macro_mesh_render.tex}
		\caption{}
	\end{subfigure}
	\caption{Fatigue specimen for macroscopic FE analysis. Geometry and boundary conditions are schematically drawn in a) and the finite element mesh with 116830 C3D8 elements is depicted in b).}
	\label{fig:macro_geom}
\end{figure}

To capture its exact loading condition and stress state, the microstructure domain is embedded in the macro model of the fatigue specimen using the Abaqus sub-modeling technique. The macro model is analyzed using a linear elastic material law with a Young's modulus of $210$\,GPa, and a Poisson's ratio of $0.3$. Figure \ref{fig:macro_geom} a) shows the geometry and boundary conditions of the specimen. On the right surface, marked in black, all degrees of freedom are fixed while the left surface, marked in blue, is cyclically loaded by a displacement amplitude of $\dispthree=0.056$\,mm along the third axis analogous to the experiment. The loading is fully reversed. Considering the static stress-strain curve, the elastic regime is not significantly exceeded by this loading, justifying the assumption of a linear elastic material for the macro model. The linear elastic material response is used as the boundary condition for the micromechanical submodel. This results in one directional coupling from the macro model to the submodel. A discretization with 116830 20-node quadratic brick elements is chosen for the specimen as shown in Figure \ref{fig:macro_geom} b). The central region for which microtexture data is available is discretized by a finer mesh, while less important areas are meshed coarsely. Three loading cycles are performed on the coupled model, at which point the cyclic FIP increment is assumed to be saturated, see for example \cite{schafer_micromechanical_2019}. 

\begin{figure}[htbp]
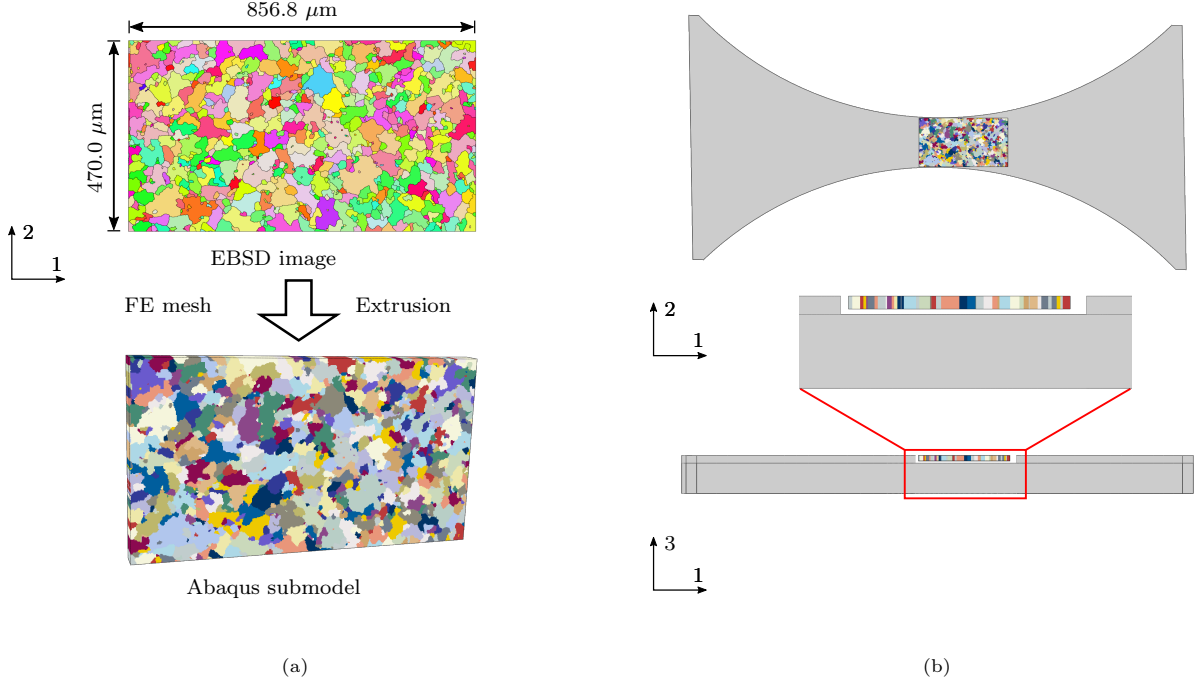

	\centering
	\begin{subfigure}{0.47\textwidth}
		\def\svgwidth{1.0\textwidth}
		\footnotesize
		\import{figures/embedding/}{ebsd2abq_render.tex}
		\caption{}
	\end{subfigure}
	\hspace{5mm}
	\begin{subfigure}{0.47\textwidth}
		\def\svgwidth{1.0\textwidth}
		\footnotesize
		\import{figures/embedding/}{embedding_render.tex}
		\caption{}
	\end{subfigure}
	\caption{Sub-modeling CPFE approach. The processing of EBSD data to an Abaqus submodel by extrusion from 2D to 3D and subsequent discretization is described in a), where the bottom represents a 3D grain unique-color representation. In b), the embedding of the microstructure within the macro geometry is shown.}
	\label{fig:embedding}
\end{figure}

As a second step, post-processing steps are performed on the distortion corrected EBSD data in MTEX, a Matlab toolbox for crystallography. Initially, data cleaning is applied using the confidence index and image quality thresholds of 0.05 and 100, respectively. Subsequently, grains are computed, setting a 5\degree{} threshold value, and all grains consisting of less than 20 indexed pixels are discarded. Finally, the EBSD data is smoothed with a spline filter, and missing points from prior operations are filled. Based on this modified EBSD data, a discretized and extruded domain for the CPFE simulation is generated, as shown in Figure \ref{fig:embedding} a). For this purpose, the grain index and orientation acquired on a hexagonal grid are transferred to a rectangular grid by nearest-neighbor interpolation. In this manner, a 2D discretized domain is set up, which is subsequently extruded to $\sfrac{\deq}{2}$ in the third dimension. Thus, the assumption of prismatic grains is made with grain boundaries (GB) extending into the direction normal to the specimen surface, which can be justified by the elongation of the grains in out-of-plane direction. The regular mesh for the investigated specimen microstructure counts $856\times 500\times 10 \approx 4.3 \times 10^6$ eight-node brick elements (C3D8). This 3D microstructure domain is embedded in the macro geometry as illustrated in Figure \ref{fig:embedding} b), where gaps in the lower illustration are for visualization purposes only. The deformation obtained from the continuum macro simulation is imposed on the CPFE model region's outer surfaces (sub-modeling).
Not shown in Figure \ref{fig:embedding} b) is an elastic embedding around the microstructure. It is $\SI{30}{\micro\meter}$ thick and used to limit the effects of load introduction.

The model is run on a high performance computing cluster node with 48GB RAM and 24 Intel(R) Xeon(R) \@ 2.60 GHz CPU.
The total runtime (wallclock time) of the macro model is about $1.9$\,h, and about $96$\,h for the microstructure model.
In the latter case, due to the relatively large number of elements for a CPFE simulation, ABAQUS's iterative solver is applied.

\subsection{Slip trace prediction}
\label{sec:slip_trace_prediction}
To assess the capability of the CPFE model to predict slip planes correctly, a comparison of simulation-based and actual slip traces, similar to \cite{Zhang2018}, is conducted. Slip trace prediction was enabled by considering the slip system exhibiting most plastic shear deformation and computing the intersection line (trace) of the respective slip plane with the surface plane as delineated in \cite{zhang2016b}. The surface intersecting line direction vector $\surfintervec$ can be determined by 
\begin{equation} \label{eq:slip_trace_line} 
	\surfintervec = \normalsurf \times (\rotmat \cdot \slipnormalmax) \, ,
\end{equation}
where $\normalsurf$ is the normal to the specimen surface on which the protrusions are observed, here it is assumed that $\normalsurf = \left(\begin{smallmatrix}0 \\ 0 \\ 1\end{smallmatrix}\right)$. $\rotmat$ denotes the crystal rotation matrix and $\slipnormalmax$ the slip plane normal.

\section{Results} \label{results}
\begin{figure}[p]
	\centering
	\begin{subfigure}{\textwidth}
	    \centering
		\def\svgwidth{\textwidth}
		\footnotesize
		\import{figures/microstructure/}{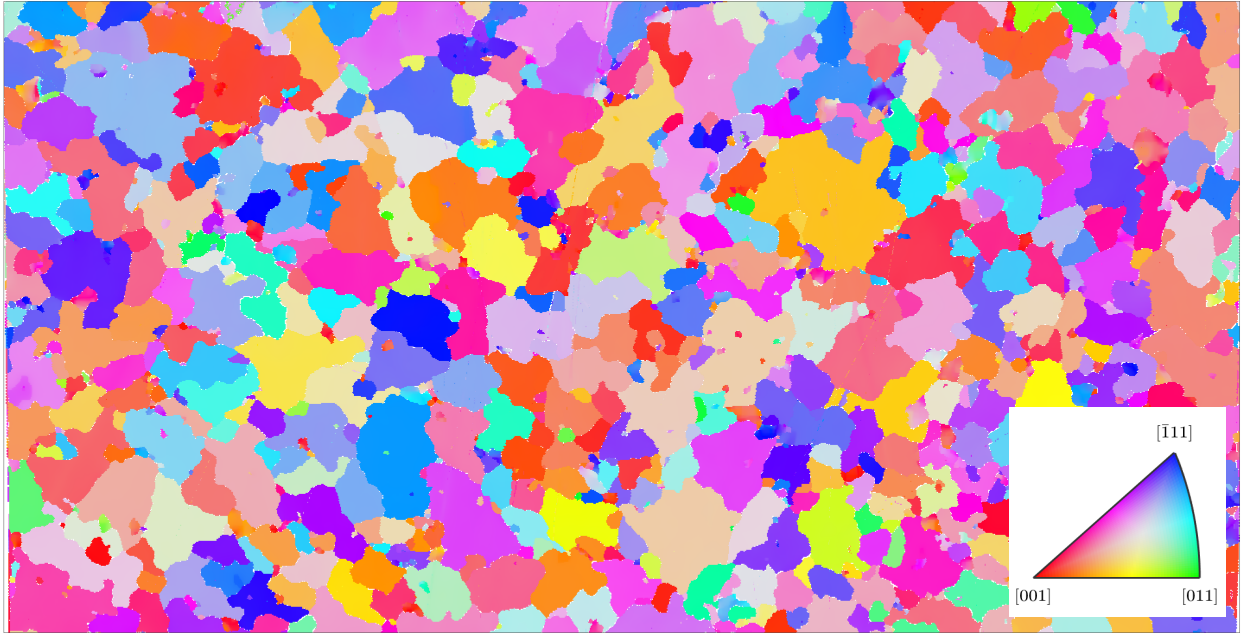}
		\caption{Reference direction: [100]}
	\end{subfigure}
	\\
	\begin{subfigure}{\textwidth}
	    \centering
		\def\svgwidth{\textwidth}
		\footnotesize
		\import{figures/microstructure/}{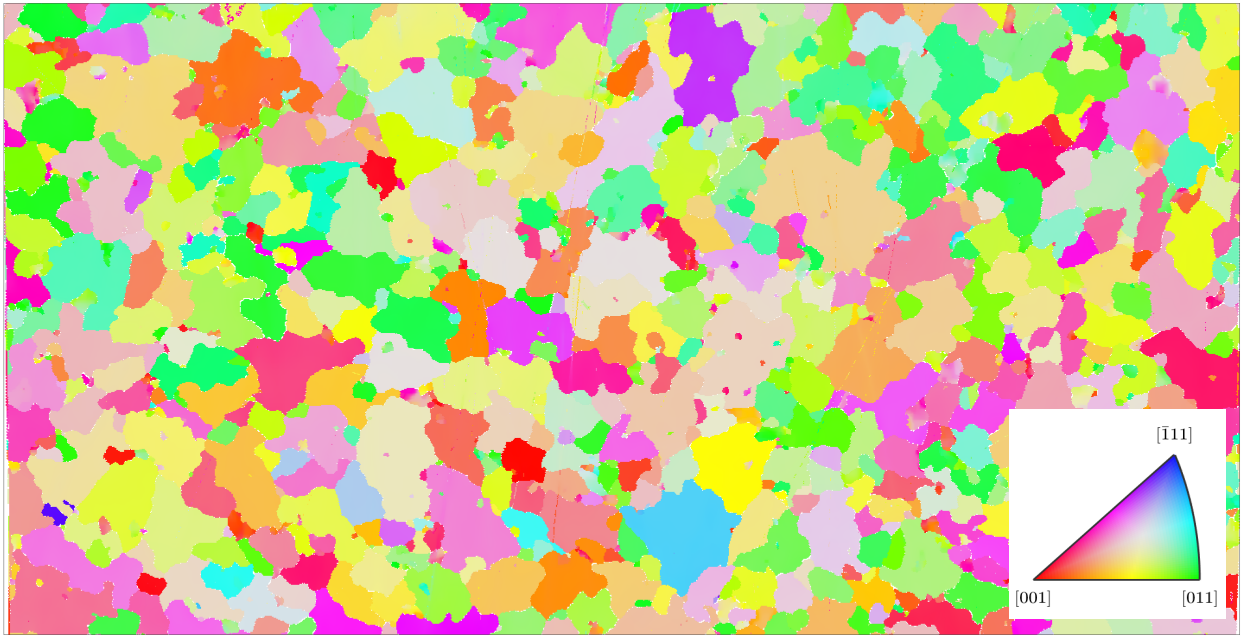}
		\caption{Reference direction: [001]}
	\end{subfigure}
	\caption{EBSD data acquired for the considered specimen for two reference directions.}
	\label{fig:ebsd}
\end{figure}

The proposed experimental and numerical workflow is applied to a mesoscale fatigue specimen. EBSD maps for the micromechanical simulation domain are given in Figure~\ref{fig:ebsd} for the sake of completeness. Experimentally acquired damage locations are presented together with calculated FIP distributions. In this work, damage locations refer to protrusions and crack sites. A qualitative slip trace analysis is performed on a selection of detected protrusions. Furthermore, one location in the microstructure where the proposed method fails to predict crack initiation is investigated in detail.

\subsection{Comparison of FIPs with experimental damage locations}
A binary map of experimental damage locations is superimposed on normalized FIP maps in Figures \ref{fig:SDV115_Overlayed} -- \ref{fig:FIPint_Overlayed} to evaluate the ability of FIP metrics to predict hot spots for damage locations reliably and to show the feasibility of the multimodal data registration and sub-modeling approach.
Figure \ref{fig:StressMises_Overlayed} shows a similar plot for the normalized von Mises stress after the last cycle.
Normalized parameters are indicated with the notation $\overline{\left(\cdot\right)}$ subsequently. Note that the damage maps superimposed on Figures \ref{fig:SDV115_Overlayed} -- \ref{fig:FIPint_Overlayed} represent the final damage stage from the segmented SEM image. This is considered as ground truth since the damage locations are predominantly spatially-confined protrusions and to a minor extent short cracks. For such damage stages the validity of the stress/deformation fields is assumed. Alternatively, individual time steps from the in-situ imaging during the fatigue experiment can be employed but are limited in terms of optical resolution.

It is worth noting that no averaging technique is applied for the plotted FIP distributions resulting in a potential mesh-dependency of the FIP solution. However, such averaging techniques (e.g. grain-based sphere averaging) tend to not impact the qualitative result, see \cite{boff_micromechanical_2016,natkowski_fatigue_2021}.


The transgranular FIP (accumulated plastic slip, Fatemi-Socie, and dissipated energy) exhibit similarities and show roughly the same number of hotspots (yellow to red regions) in the same grains. A subset of the most pronounced hotspots where the FIP exceeds 80\% of its maximum value are annotated with red arrows for better visibility, in Figures \ref{fig:SDV115_Overlayed} -- \ref{fig:FIPint_Overlayed}. Distinct discontinuities are present in each FIP distribution in the highest loaded section of the specimen. As evident, damage locations indicated in black often arise at such discontinuities but not necessarily in the grains exhibiting the highest FIP values. Overall, from comparison to the grain maps in Figure \ref{fig:ebsd} it can be inferred that transgranular FIP distributions vary within grains and tend to localize predominantly at grain boundaries and at triple points. This is in agreement with previous investigations, e.g., by \cite{basseville_numerical_2017}, or \cite{el_shawish_combining_2017}.
The intergranular FIP is only defined at the grain boundaries themselves and is displayed in Figure \ref{fig:FIPint_Overlayed}. High values in the intergranular FIP often spatially coincide with hotspots in the other FIPs.

Ideally, the critical red areas, where the simulation predicts the highest FIP, should follow the experimentally observed damage locations. However, the majority of damage locations are located outside of highest FIP regions but rather in regions of moderately high FIP (green regions). Some of regions where FIP and ground truth are not aligned are annotated in Figure \ref{fig:SDV115_Overlayed} with box annotations exemplarily. These regions are explored subsequently.


\tikzset{align at top/.style={baseline=(current bounding box.north)}}
\newcommand{\hflip}[1]{\reflectbox{\rotatebox[origin=c]{180}{#1}}}
\begin{figure}[p]
	\centering
	\footnotesize
	\hflip{
\begin{tikzpicture}

\begin{axis}[
axis equal image,
width=1.1\textwidth,
hide x axis,
hide y axis,
tick align=outside,
tick pos=left,
x grid style={white!69.0196078431373!black},
xmin=0, xmax=855.6,
xtick style={color=black},
y grid style={white!69.0196078431373!black},
ymin=30.484095, ymax=499.084095,
ytick style={color=black}
]
\addplot graphics [includegraphics cmd=\pgfimage,xmin=0, xmax=856.20003, ymin=30.484095, ymax=500.401132046717] {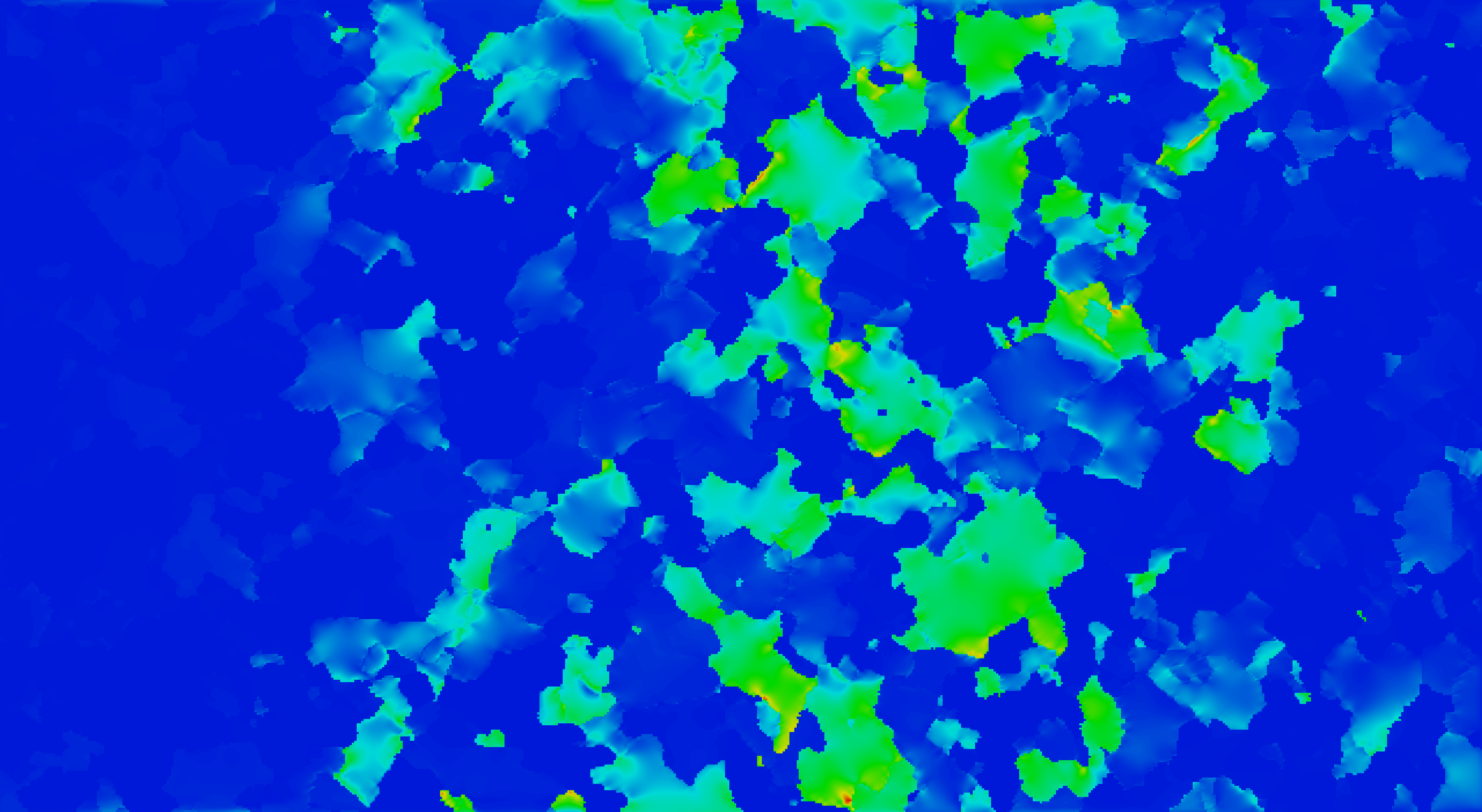};
\addplot[opacity=0.5] graphics [includegraphics cmd=\pgfimage,xmin=0, xmax=855.6, ymin=30.484095, ymax=499.084095] {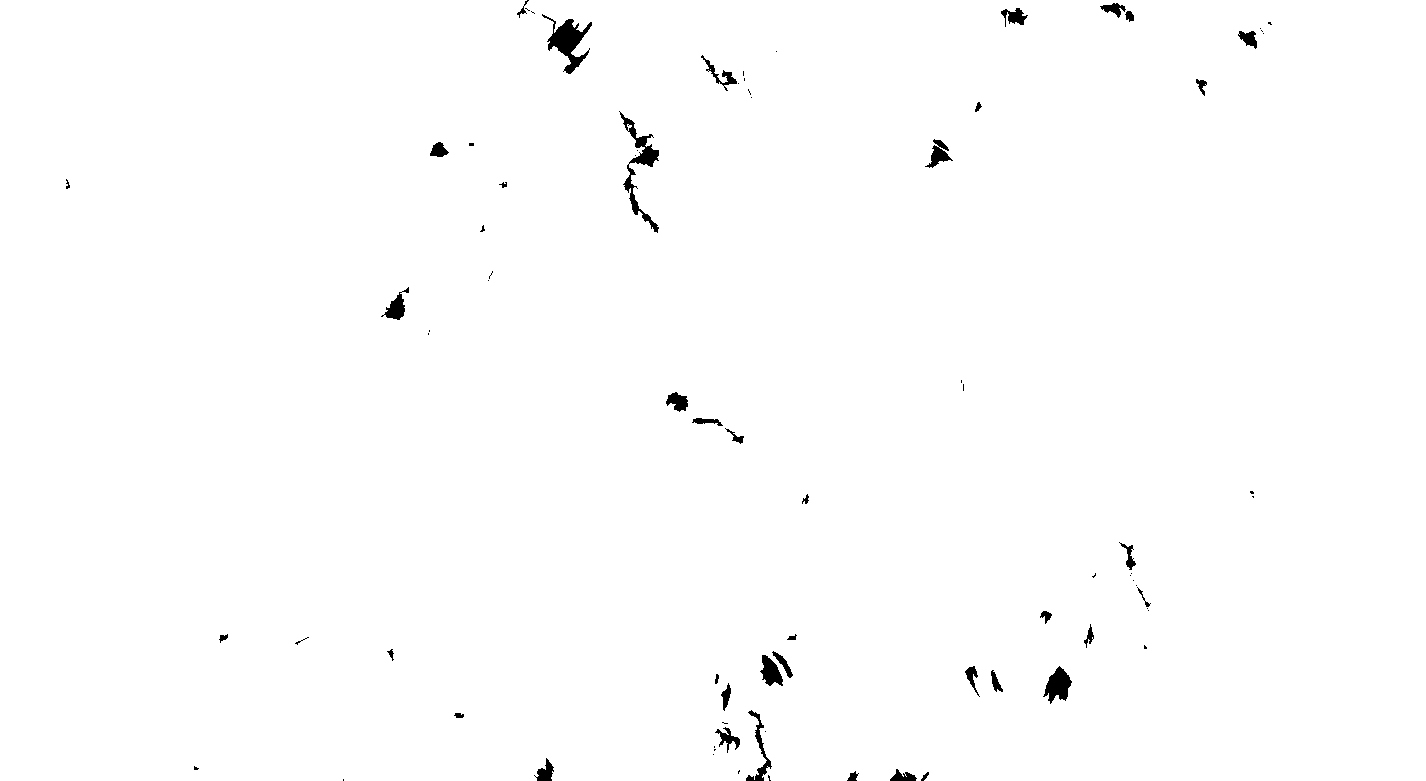};
\draw [line width=0.5mm, white ] (320,425) rectangle (360,465);
\draw [line width=0.5mm, white ] (443,47) rectangle (487,90);
\draw [line width=0.5mm, black ] (660,92) rectangle (705,160);
\draw[->,line width=0.5mm, red](637.1039652824402,298.30286882045743)--(647.1039652824402,288.30286882045743);
\draw[->,line width=0.5mm, red](595.5803000926971,105.68403682830811)--(605.5803000926971,95.68403682830811);
\draw[->,line width=0.5mm, red](592.9013395309448,468.407819486885)--(602.9013395309448,458.407819486885);
\draw[->,line width=0.5mm, red](558.075031042099,108.18557671430588)--(568.075031042099,98.18557671430588);
\draw[->,line width=0.5mm, red](519.2302668094635,437.13853738430015)--(529.2302668094635,427.13853738430015);
\draw[->,line width=0.5mm, red](503.15659284591675,440.89084348800657)--(513.1565928459167,430.89084348800657);
\draw[->,line width=0.5mm, red](497.7987015247345,215.751940823822)--(507.7987015247345,205.751940823822);
\draw[->,line width=0.5mm, red](487.0829039812088,434.6369900477218)--(497.0829039812088,424.6369900477218);
\draw[->,line width=0.5mm, red](483.0644929409027,11.876156992940903)--(493.0644929409027,1.8761569929409028);
\draw[->,line width=0.5mm, red](476.3671213388443,262.03049204471586)--(486.3671213388443,252.03049204471588);
\draw[->,line width=0.5mm, red](473.6881756782532,279.54129359844205)--(483.6881756782532,269.54129359844205);
\draw[->,line width=0.5mm, red](452.256595492363,88.17322782400133)--(462.256595492363,78.17322782400133);
\draw[->,line width=0.5mm, red](449.5776498317719,63.15779543641091)--(459.5776498317719,53.15779543641091);
\draw[->,line width=0.5mm, red](440.2013325691223,44.39622021439553)--(450.2013325691223,34.39622021439553);
\draw[->,line width=0.5mm, red](436.18290662765503,74.41473982456208)--(446.18290662765503,64.41473982456208);
\draw[->,line width=0.5mm, red](430.8250153064728,378.3522643816757)--(440.8250153064728,368.3522643816757);
\draw[->,line width=0.5mm, red](389.3013352155685,460.90317747715)--(399.3013352155685,450.90317747715);
\draw[->,line width=0.5mm, red](323.6671143770218,18.13001602116108)--(333.6671143770218,8.13001602116108);
\end{axis}

\end{tikzpicture}} \\
\begin{tikzpicture}

\begin{axis}[
width=6.5cm,
height=2cm,
axis line style ={white!69.0196078431373!white},
tick align=outside,
tick pos=left,
x grid style={white!69.0196078431373!white},
xmin=-0.5, xmax=12.5,
xtick style={color=black},
xtick={-0.5,3.0,6.5},
xticklabels={$\SI{0}{}$,$\fippover$,$\SI{1}{}$},
y dir=reverse,
y grid style={white!69.0196078431373!white},
ymin=-0.5, ymax=0.5,
ytick style={color=white},
axis y line=none,
y grid style={white!69.0196078431373!white},
ymin=-0.5, ymax=0.5,
]
\addplot graphics [includegraphics cmd=\pgfimage,xmin=-0.5, xmax=6.5, ymin=0.5, ymax=-0.5] {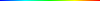};
\node[fill=black!40!black,
      anchor=north west,
      thick,
      draw,
      minimum height=0.25,
      minimum width=0.25,
      label=0:Damage,
] at (axis cs: 8,-0.25) {};
\end{axis}

\end{tikzpicture} 
	\caption{$\fippover$ distribution superimposed with damage locations colored in black. The white boxes on the bottom and the top (left and right) side mark the damage spots shown in the subsequent slip trace analysis SEM images of Figure \ref{fig:slip_traces}. The additional black box annotates the region that is subsequently considered in the cyclic damage evolution analysis to investigate the inadequacies of the model. The red arrows indicate FIP values which are greater than 80\% of the maximum FIP value found with a search radius of 10 element edge lengths.}
	\label{fig:SDV115_Overlayed}
\end{figure}
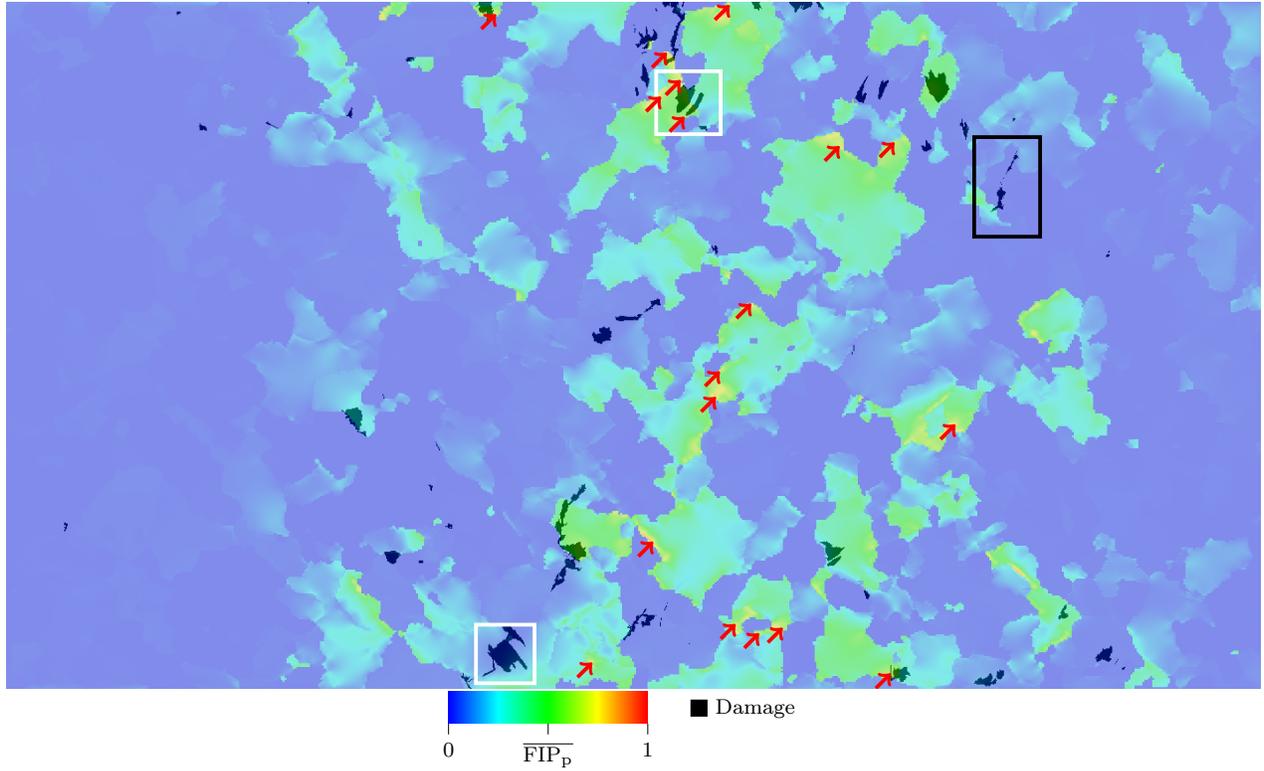

\begin{figure}[p]
	\centering
	\footnotesize
	\hflip{
\begin{tikzpicture}

\begin{axis}[
axis equal image,
width=1.1\textwidth,
hide x axis,
hide y axis,
tick align=outside,
tick pos=left,
x grid style={white!69.0196078431373!black},
xmin=0, xmax=855.6,
xtick style={color=black},
y grid style={white!69.0196078431373!black},
ymin=30.484095, ymax=499.084095,
ytick style={color=black}
]
\addplot graphics [includegraphics cmd=\pgfimage,xmin=0, xmax=856.20003, ymin=30.484095, ymax=500.401132046717] {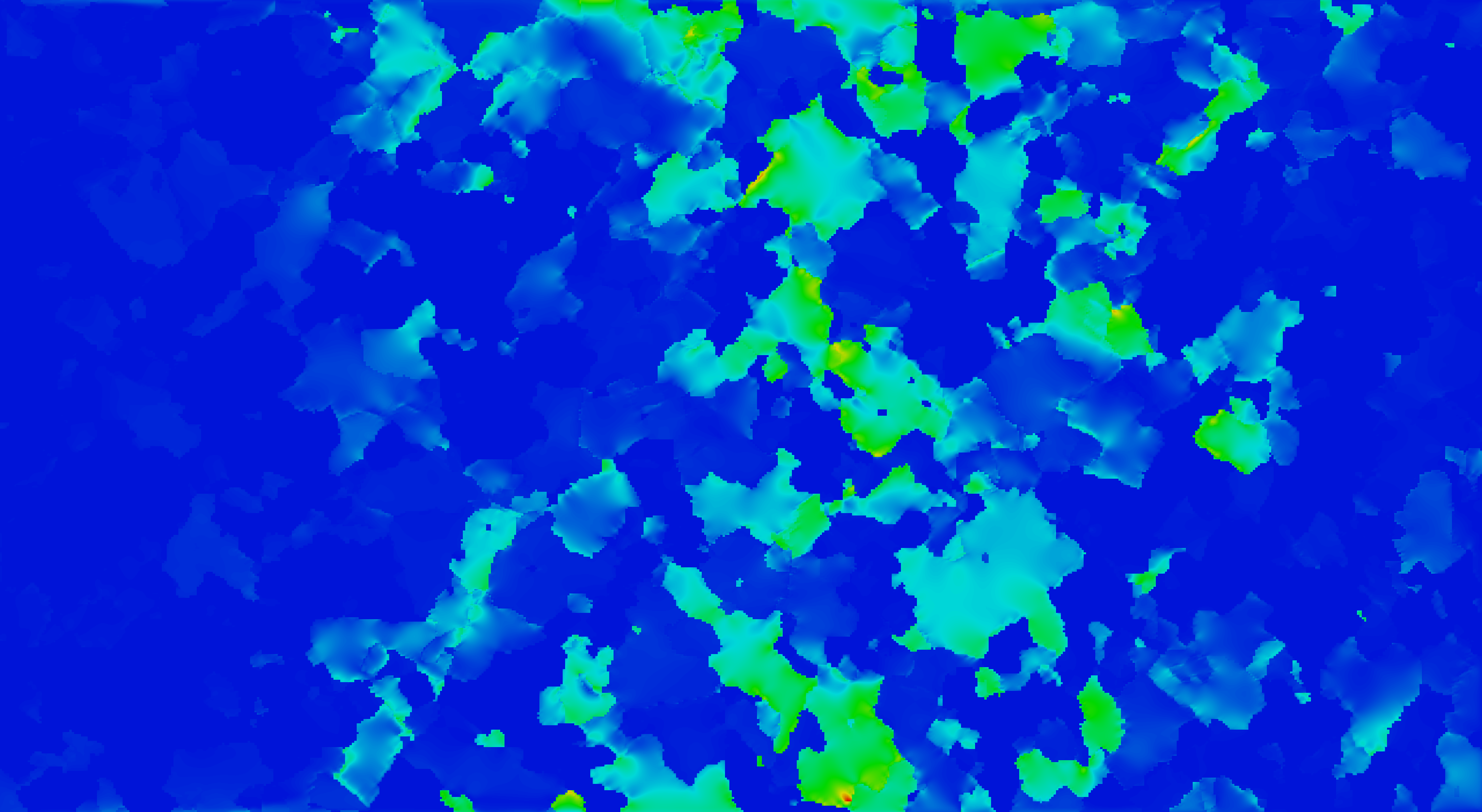};
\addplot[opacity=0.5] graphics [includegraphics cmd=\pgfimage,xmin=0, xmax=855.6, ymin=30.484095, ymax=499.084095] {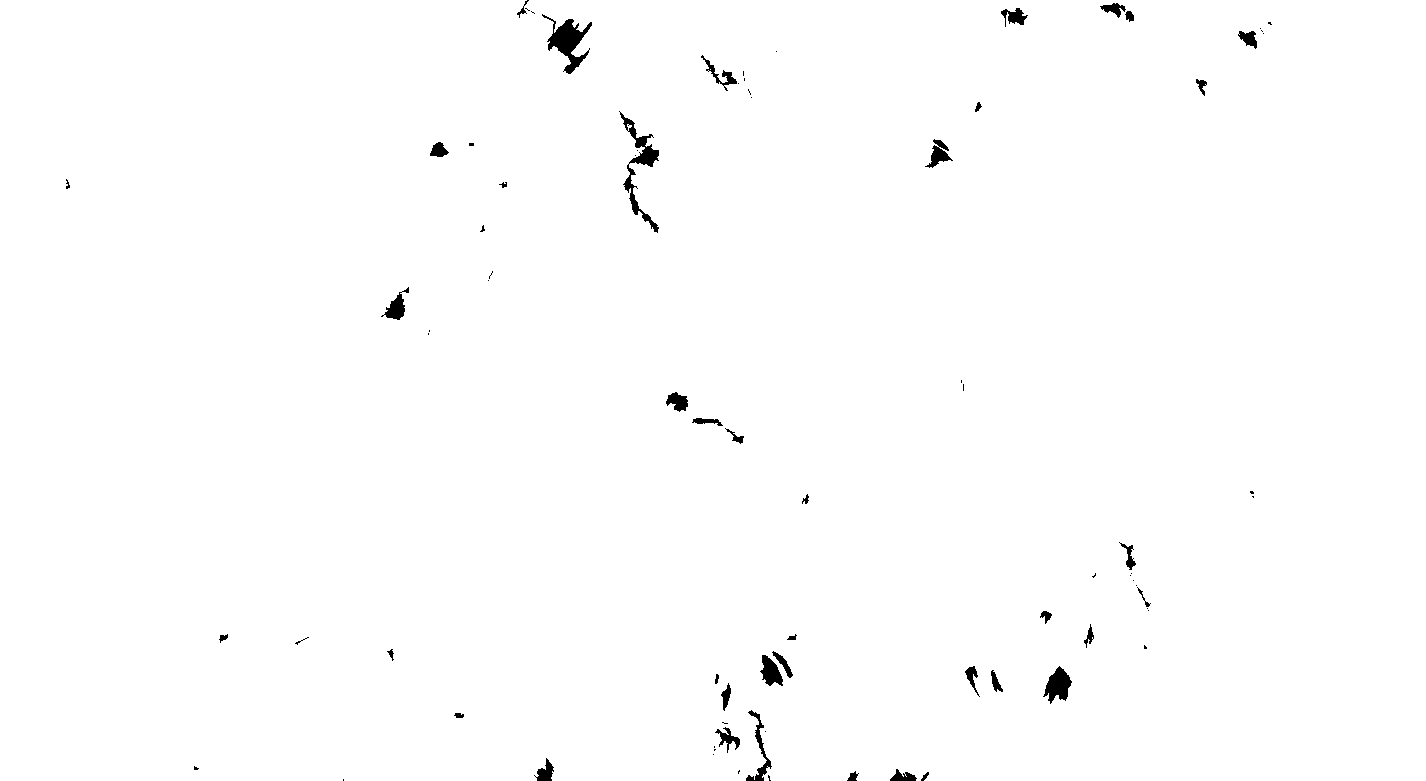};
\draw[->,line width=0.5mm, red](635.7644999027252,298.30286882045743)--(645.7644999027252,288.30286882045743);
\draw[->,line width=0.5mm, red](481.7250126600266,16.879244215517037)--(491.7250126600266,6.879244215517037);
\draw[->,line width=0.5mm, red](390.6408005952835,458.40165994289396)--(400.6408005952835,448.40165994289396);
\draw[->,line width=0.5mm, red](322.3276489973068,19.380787826805122)--(332.3276489973068,9.380787826805122);
\end{axis}

\end{tikzpicture}}  \\
\begin{tikzpicture}

\begin{axis}[
width=6.5cm,
height=2cm,
axis line style ={white!69.0196078431373!white},
tick align=outside,
tick pos=left,
x grid style={white!69.0196078431373!white},
xmin=-0.5, xmax=12.5,
xtick style={color=black},
xtick={-0.5,3.0,6.5},
xticklabels={$\SI{0}{}$,$\fipfsover$,$\SI{1}{}$},
y dir=reverse,
y grid style={white!69.0196078431373!white},
ymin=-0.5, ymax=0.5,
ytick style={color=white},
axis y line=none,
y grid style={white!69.0196078431373!white},
ymin=-0.5, ymax=0.5,
]
\addplot graphics [includegraphics cmd=\pgfimage,xmin=-0.5, xmax=6.5, ymin=0.5, ymax=-0.5] {figures/resultsnew/legend-fip.png};
\node[fill=black!40!black,
      anchor=north west,
      thick,
      draw,
      minimum height=0.25,
      minimum width=0.25,
      label=0:Damage,
] at (axis cs: 8,-0.25) {};
\end{axis}

\end{tikzpicture} \\
	\caption{$\fipfsover$ distribution superimposed with damage locations, like in Figure \ref{fig:SDV115_Overlayed}.}
	\label{fig:SDV116_Overlayed}
\end{figure}

\begin{figure}[p]
	\centering
	\footnotesize
	\hflip{
\begin{tikzpicture}

\begin{axis}[
axis equal image,
width=1.1\textwidth,
hide x axis,
hide y axis,
tick align=outside,
tick pos=left,
x grid style={white!69.0196078431373!black},
xmin=0, xmax=855.6,
xtick style={color=black},
y grid style={white!69.0196078431373!black},
ymin=30.484095, ymax=499.084095,
ytick style={color=black}
]
\addplot graphics [includegraphics cmd=\pgfimage,xmin=0, xmax=856.20003, ymin=30.484095, ymax=500.401132046717] {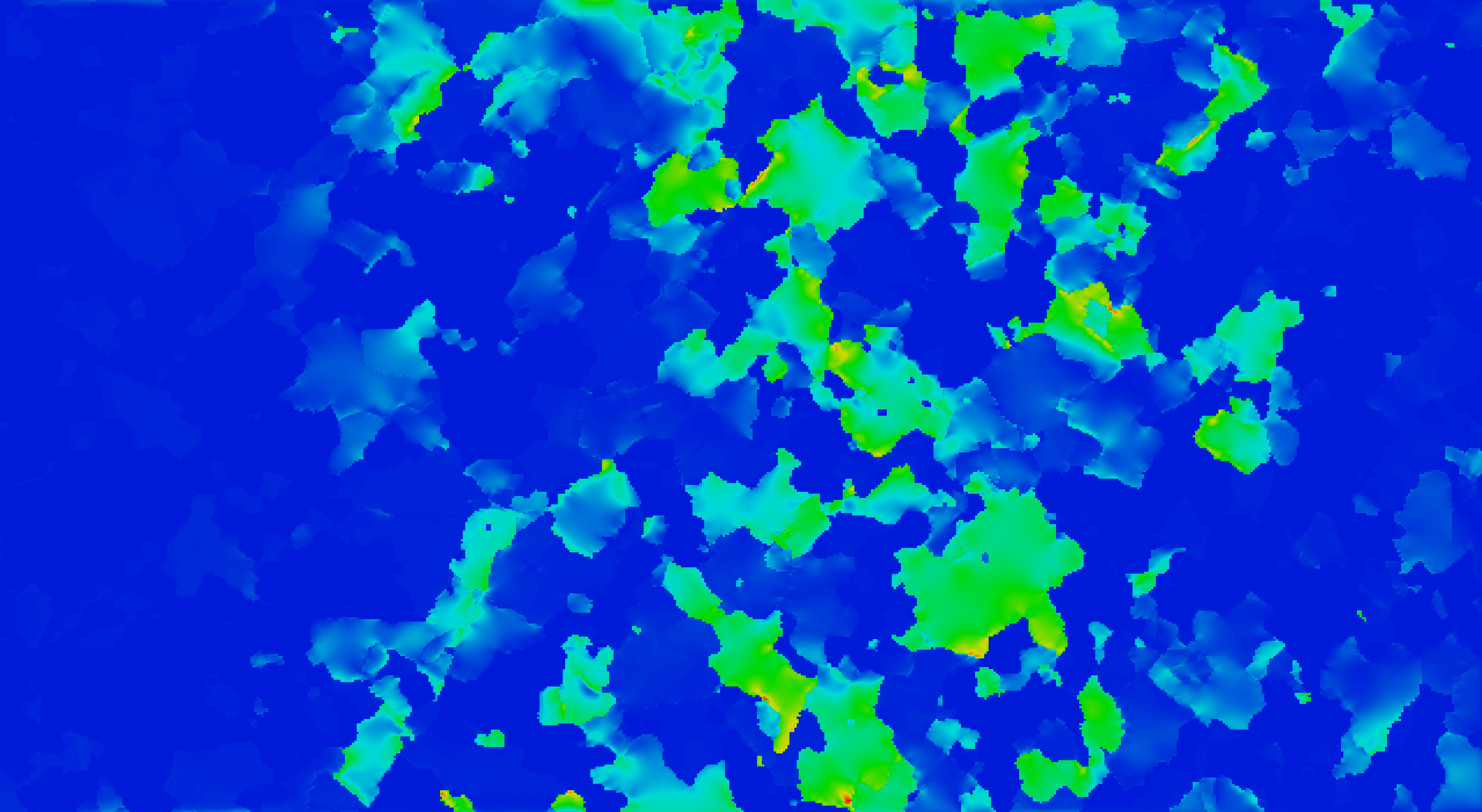};
\addplot[opacity=0.5] graphics [includegraphics cmd=\pgfimage,xmin=0, xmax=855.6, ymin=30.484095, ymax=499.084095] {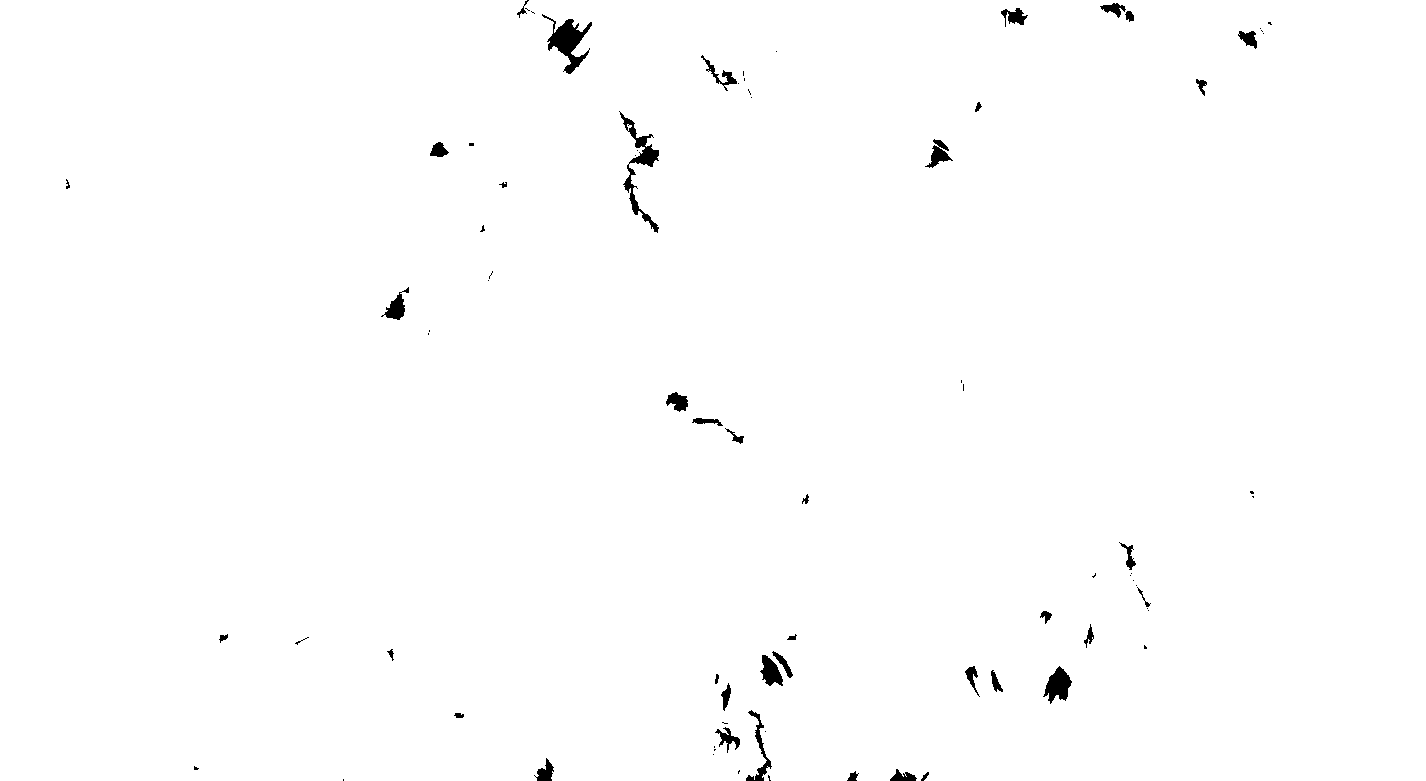};
\draw[->,line width=0.5mm, red](635.7644999027252,298.30286882045743)--(645.7644999027252,288.30286882045743);
\draw[->,line width=0.5mm, red](595.5803000926971,105.68403682830811)--(605.5803000926971,95.68403682830811);
\draw[->,line width=0.5mm, red](559.414496421814,108.18557671430588)--(569.414496421814,98.18557671430588);
\draw[->,line width=0.5mm, red](517.8908014297485,437.13853738430015)--(527.8908014297485,427.13853738430015);
\draw[->,line width=0.5mm, red](503.15659284591675,440.89084348800657)--(513.1565928459167,430.89084348800657);
\draw[->,line width=0.5mm, red](483.0644929409027,11.876156992940903)--(493.0644929409027,1.8761569929409028);
\draw[->,line width=0.5mm, red](449.5776498317719,63.15779543641091)--(459.5776498317719,53.15779543641091);
\draw[->,line width=0.5mm, red](436.18290662765503,74.41473982456208)--(446.18290662765503,64.41473982456208);
\draw[->,line width=0.5mm, red](412.0723807811737,359.59068915966026)--(422.0723807811737,349.59068915966026);
\draw[->,line width=0.5mm, red](389.3013352155685,459.652418710022)--(399.3013352155685,449.652418710022);
\draw[->,line width=0.5mm, red](322.3276489973068,19.380787826805122)--(332.3276489973068,9.380787826805122);
\draw[->,line width=0.5mm, red](245.97764551639563,20.63155963244915)--(255.97764551639563,10.63155963244915);
\end{axis}

\end{tikzpicture}}  \\
\begin{tikzpicture}

\begin{axis}[
width=6.5cm,
height=2cm,
axis line style ={white!69.0196078431373!white},
tick align=outside,
tick pos=left,
x grid style={white!69.0196078431373!white},
xmin=-0.5, xmax=12.5,
xtick style={color=black},
xtick={-0.5,3.0,6.5},
xticklabels={$\SI{0}{}$,$\fipwover$,$\SI{1}{}$},
y dir=reverse,
y grid style={white!69.0196078431373!white},
ymin=-0.5, ymax=0.5,
ytick style={color=white},
axis y line=none,
y grid style={white!69.0196078431373!white},
ymin=-0.5, ymax=0.5,
]
\addplot graphics [includegraphics cmd=\pgfimage,xmin=-0.5, xmax=6.5, ymin=0.5, ymax=-0.5] {figures/resultsnew/legend-fip.png};
\node[fill=black!40!black,
      anchor=north west,
      thick,
      draw,
      minimum height=0.25,
      minimum width=0.25,
      label=0:Damage,
] at (axis cs: 8,-0.25) {};
\end{axis}

\end{tikzpicture} \\
	\caption{$\fipwover$ distribution superimposed with damage locations, like in Figure \ref{fig:SDV115_Overlayed}.}
	\label{fig:SDV183_Overlayed}
\end{figure}

\begin{figure}[p]
	\centering
	\footnotesize
	\hflip{
\begin{tikzpicture}

\begin{axis}[
axis equal image,
width=1.1\textwidth,
hide x axis,
hide y axis,
tick align=outside,
tick pos=left,
x grid style={white!69.0196078431373!black},
xmin=0, xmax=855.6,
xtick style={color=black},
y grid style={white!69.0196078431373!black},
ymin=30.484095, ymax=499.084095,
ytick style={color=black}
]
\addplot graphics [includegraphics cmd=\pgfimage,xmin=0, xmax=856.20003, ymin=30.484095, ymax=499.084095] {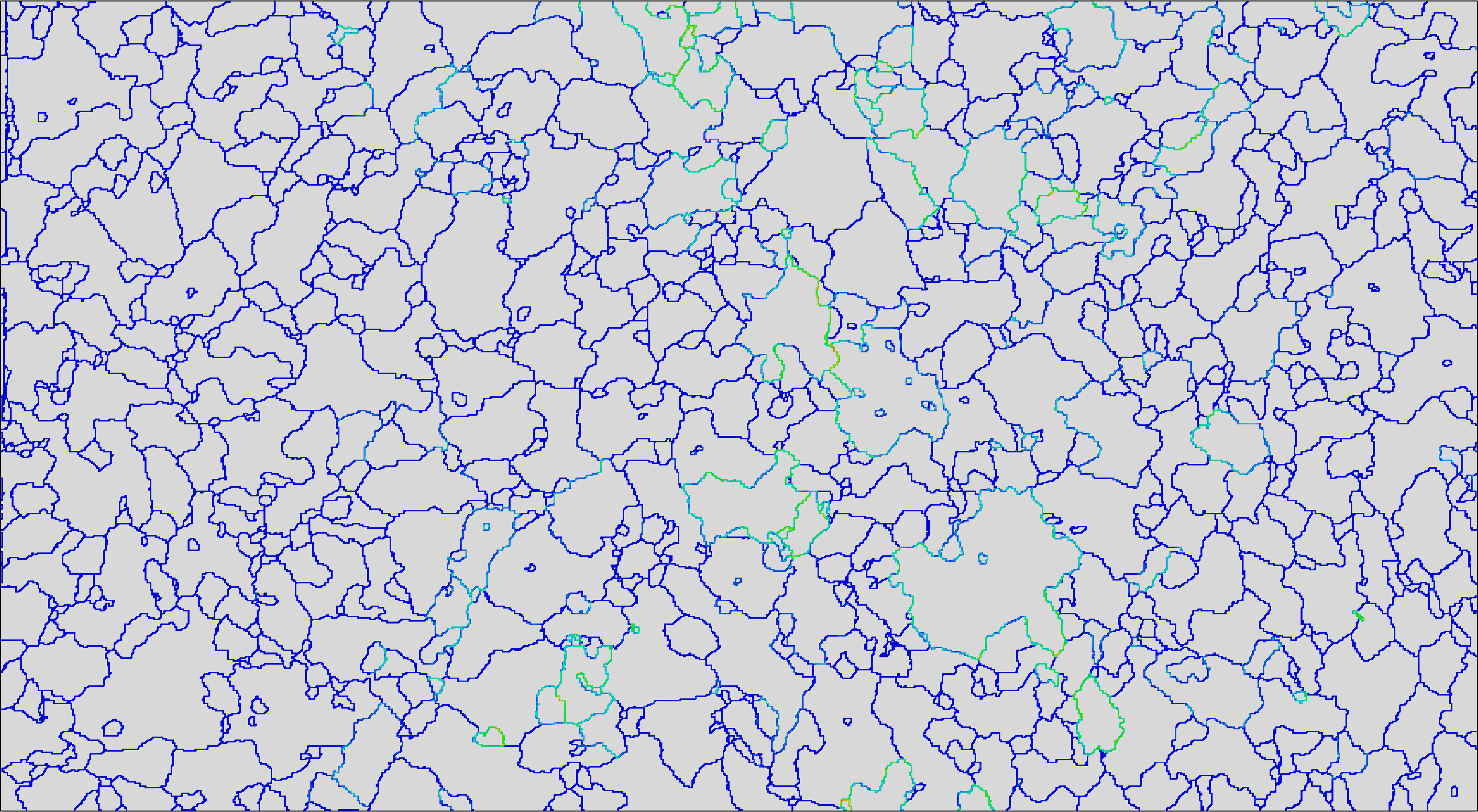};
\addplot[opacity=0.5] graphics [includegraphics cmd=\pgfimage,xmin=0, xmax=855.6, ymin=30.484095, ymax=499.084095] {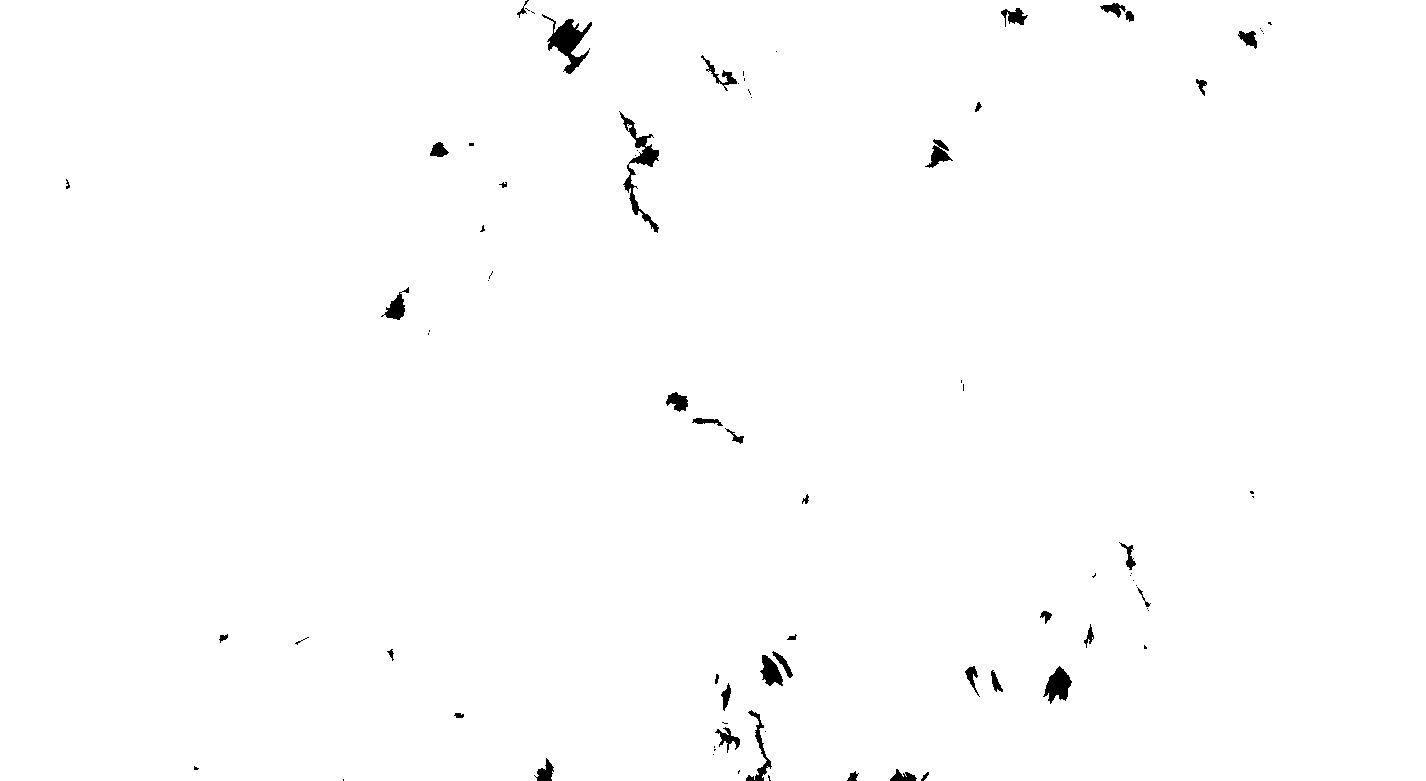};
\draw[->,line width=0.5mm, red](695.3710818290709,242.6435299885559)--(705.3710818290709,232.6435299885559);
\draw[->,line width=0.5mm, red](677.9579424858093,396.4884527695465)--(687.9579424858093,386.4884527695465);
\draw[->,line width=0.5mm, red](471.6789627075195,280.1666804325867)--(481.6789627075195,270.1666804325867);
\draw[->,line width=0.5mm, red](475.6973886489868,261.40510521057126)--(485.6973886489868,251.4051052105713);
\draw[->,line width=0.5mm, red](464.9816060066223,316.4390572083282)--(474.9816060066223,306.4390572083282);
\end{axis}
\end{tikzpicture}}  \\
\begin{tikzpicture}

\begin{axis}[
width=6.5cm,
height=2cm,
axis line style ={white!69.0196078431373!white},
tick align=outside,
tick pos=left,
x grid style={white!69.0196078431373!white},
xmin=-0.5, xmax=12.5,
xtick style={color=black},
xtick={-0.5,3.0,6.5},
xticklabels={$\SI{0}{}$,$\fipintover$,$\SI{1}{}$},
y dir=reverse,
y grid style={white!69.0196078431373!white},
ymin=-0.5, ymax=0.5,
ytick style={color=white},
axis y line=none,
y grid style={white!69.0196078431373!white},
ymin=-0.5, ymax=0.5,
]
\addplot graphics [includegraphics cmd=\pgfimage,xmin=-0.5, xmax=6.5, ymin=0.5, ymax=-0.5] {figures/resultsnew/legend-fip.png};
\node[fill=black!40!black,
      anchor=north west,
      thick,
      draw,
      minimum height=0.25,
      minimum width=0.25,
      label=0:Damage,
] at (axis cs: 8,-0.25) {};
\end{axis}

\end{tikzpicture} \\
	\caption{$\fipintover$ distribution superimposed with damage locations, like in Figure \ref{fig:SDV115_Overlayed}.}
	\label{fig:FIPint_Overlayed}
\end{figure}

\begin{figure}[p]
	\centering
	\footnotesize
	\hflip{
\begin{tikzpicture}

\begin{axis}[
axis equal image,
width=1.1\textwidth,
hide x axis,
hide y axis,
tick align=outside,
tick pos=left,
x grid style={white!69.0196078431373!black},
xmin=0, xmax=855.6,
xtick style={color=black},
y grid style={white!69.0196078431373!black},
ymin=30.484095, ymax=499.084095,
ytick style={color=black}
]
\addplot graphics [includegraphics cmd=\pgfimage,xmin=0, xmax=856.20003, ymin=30.484095, ymax=500.708367768879] {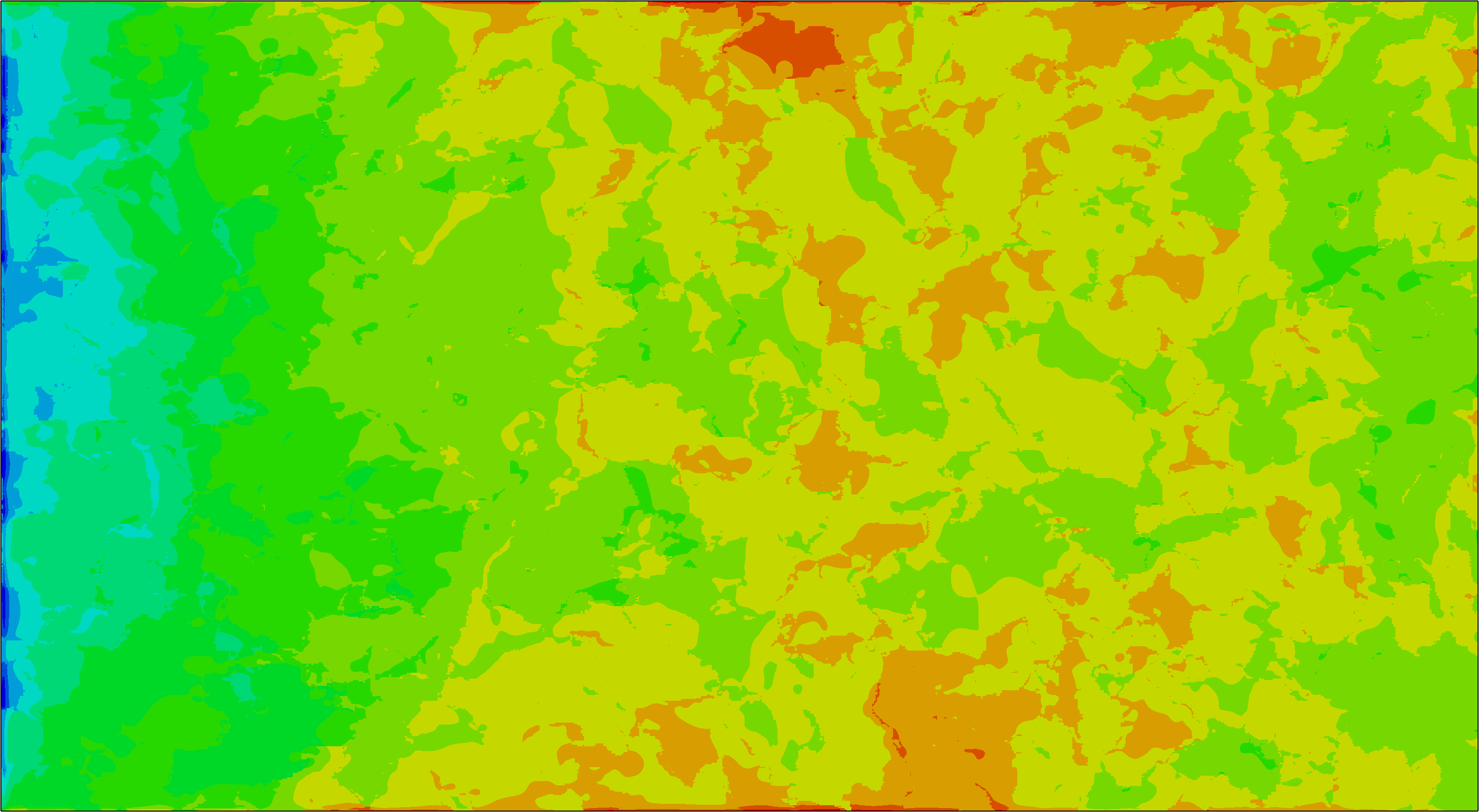};
\addplot[opacity=0.5] graphics [includegraphics cmd=\pgfimage,xmin=0, xmax=855.6, ymin=30.484095, ymax=499.084095] {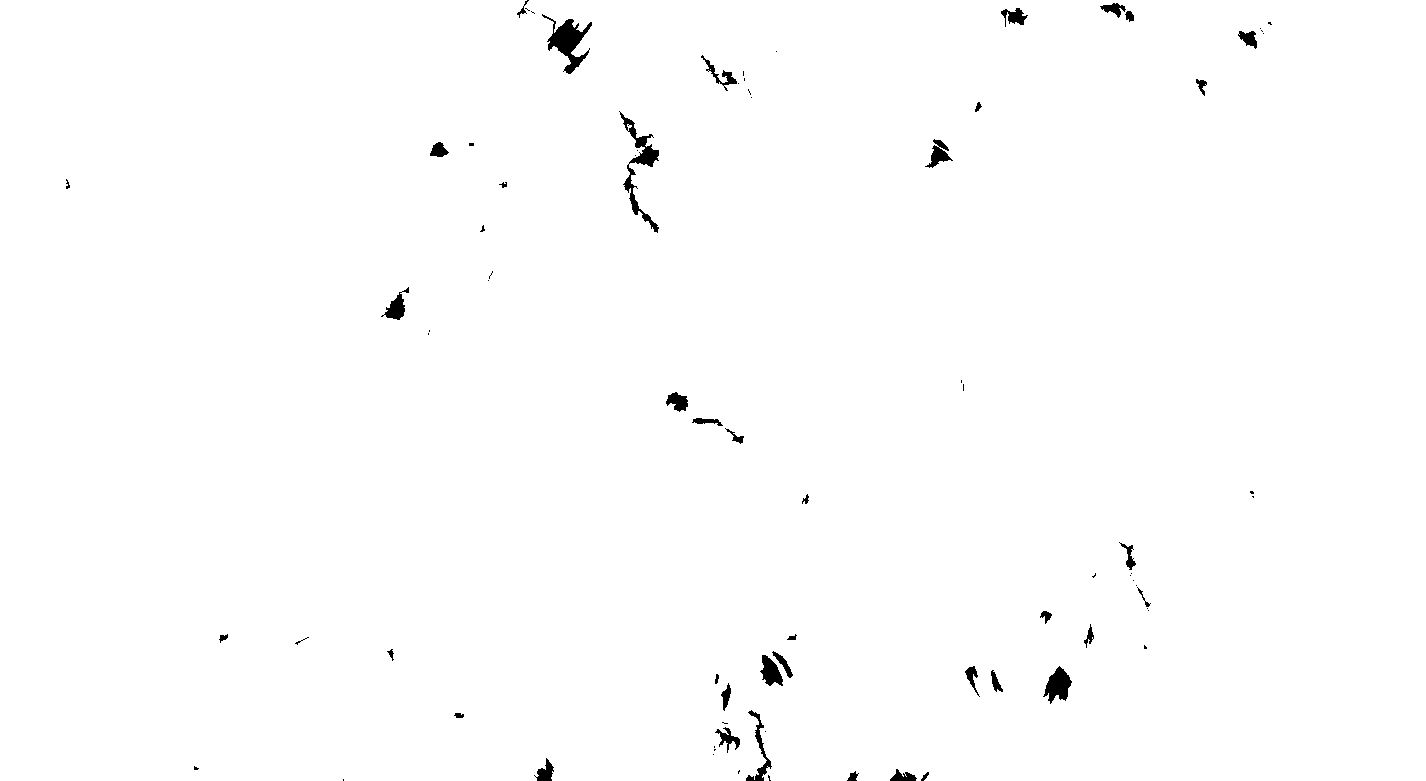};
\end{axis}

\end{tikzpicture}}  \\
\begin{tikzpicture}

\begin{axis}[
width=6.5cm,
height=2cm,
axis line style ={white!69.0196078431373!white},
tick align=outside,
tick pos=left,
x grid style={white!69.0196078431373!white},
xmin=-0.5, xmax=12.5,
xtick style={color=black},
xtick={-0.5,3.0,6.5},
xticklabels={$\SI{0}{}$,$\stressmisesover$,$\SI{1}{}$},
y dir=reverse,
y grid style={white!69.0196078431373!white},
ymin=-0.5, ymax=0.5,
ytick style={color=white},
axis y line=none,
y grid style={white!69.0196078431373!white},
ymin=-0.5, ymax=0.5,
]
\addplot graphics [includegraphics cmd=\pgfimage,xmin=-0.5, xmax=6.5, ymin=0.5, ymax=-0.5] {figures/resultsnew/legend-fip.png};
\node[fill=black!40!black,
      anchor=north west,
      thick,
      draw,
      minimum height=0.25,
      minimum width=0.25,
      label=0:Damage,
] at (axis cs: 8,-0.25) {};
\end{axis}

\end{tikzpicture} \\
	\caption{$\stressmisesover$ distribution superimposed with damage locations, like in Figure \ref{fig:SDV115_Overlayed}. Arrows indicating high values are not shown.}
	\label{fig:StressMises_Overlayed}
\end{figure}

\subsection{Slip trace analysis}
\begin{figure}[htbp]
	\centering
	\footnotesize
	\adjustbox{width=0.9\textwidth,keepaspectratio}{\import{figures/resultsnew/}{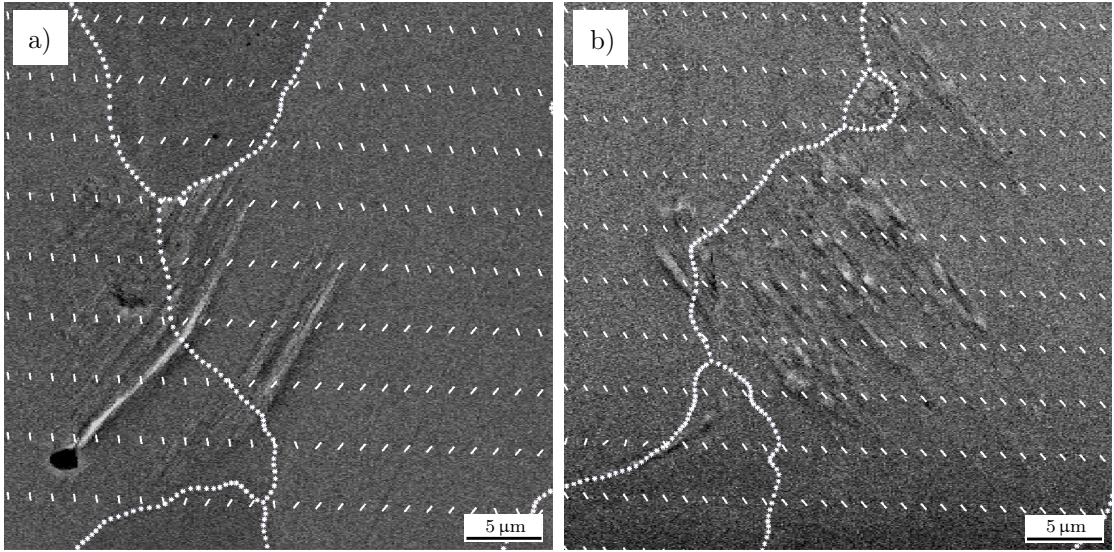}} 
	\caption{Comparison of protrusion slip trace topography with the  simulation-derived, element-wise trace of the slip plane containing the most activated slip system (white lines). The two instances illustrated here are marked by white boxes in Figure \ref{fig:SDV115_Overlayed}. White dotted lines show the grain boundaries derived from electron backscatter diffraction}
	\label{fig:slip_traces}
\end{figure}

The qualitative comparison of eight protrusions and their slip trace orientations with predicted slip trace orientations (according to Section \ref{sec:slip_trace_prediction}) shows good agreement. A subset of two protrusions is shown in Figure \ref{fig:slip_traces}. The white lines indicate the predicted slip traces per finite element. Angular deviations between protrusion traces and predicted traces are typically in the order of a few degrees.

The dotted white lines indicate grain boundaries inferred from EBSD data. The characteristic slip traces within the protrusion in Figure \ref{fig:slip_traces} a) undergo a direction change when transitioning across the grain boundary. While the prediction in the right grain is aligned with the observed protrusion, there is a deviation observed for the left grain. This discrepancy can be ascribed to two potential reasons. One possibility is that the pore resulting from a prior manganese sulfide inclusion causes an alteration of the stress state, which is not considered in the modeling. Another option would be that the protrusion emerges in the right grain and then upon transition undergoes a direction change towards the closest aligned slip system rather than the individually highest loaded slip system in the left grain. Interestingly, while in most grains, the predicted slip traces do not vary within grains, in the right grain, the prediction is correct only in the vicinity of the actual damage location. The predicted slip trace in Figure \ref{fig:slip_traces} b) is in accordance with the visible traces in the protrusion. Aside from the shown predominant mode of single slip, few protrusions with comparatively more intricate topographies indicating multiple slip are observable in other regions. 

\subsection{Cyclic evolution of crack formation and growth}
Since some damage locations occurred in regions where lower FIP metrics were predicted, the question arises which shortcoming leads to this discrepancy. Therefore, such a damage location was investigated in detail. Figure \ref{fig:time_series} depicts an image series of in-situ light optical images signifying damage (dark grayscale values) overlayed with a grain map or the accumulated plasticity FIP. The overlayed damage maps are difference images where a initial reference image was subtracted from an image associated with the specified cycle. The crack nucleates in the vicinity of a fabrication-induced pore present at the grain boundary. Crack growth occurs in both directions consecutively, indicated by the arrows in Figure \ref{fig:time_series} b). Finally, grain boundaries pose a barrier to both crack branches. Therefore, damage locations to a minor extent can be ascribed to pore-induced stress concentration. Pores present at the surface exhibit equivalent diameters of 0.1--6.0\,$\upmu$m.

\newcommand{\specialcell}[2][c]{%
  \begin{tabular}[#1]{@{}c@{}}#2\end{tabular}}
\newcommand{\tikzscale}[1]{\pgfplotsset{every axis/.append style={scale = #1}}}

\begin{figure}[htbp]
	\centering
	\begin{subfigure}{0.65\textwidth}
		\footnotesize
        \tikzscale{0.5}
        \begin{tabularx}{0.65\textwidth}{llll}
        \specialcell{\hflip{
\begin{tikzpicture}

\begin{axis}[
axis equal image,
hide x axis,
hide y axis,
tick align=outside,
tick pos=left,
x grid style={white!69.0196078431373!black},
xmin=660, xmax=720,
xtick style={color=black},
y grid style={white!69.0196078431373!black},
ymin=118, ymax=200,
ytick style={color=black}
]
\addplot graphics [includegraphics cmd=\pgfimage,xmin=0, xmax=856.20003, ymin=30.484095, ymax=500.708367768879] {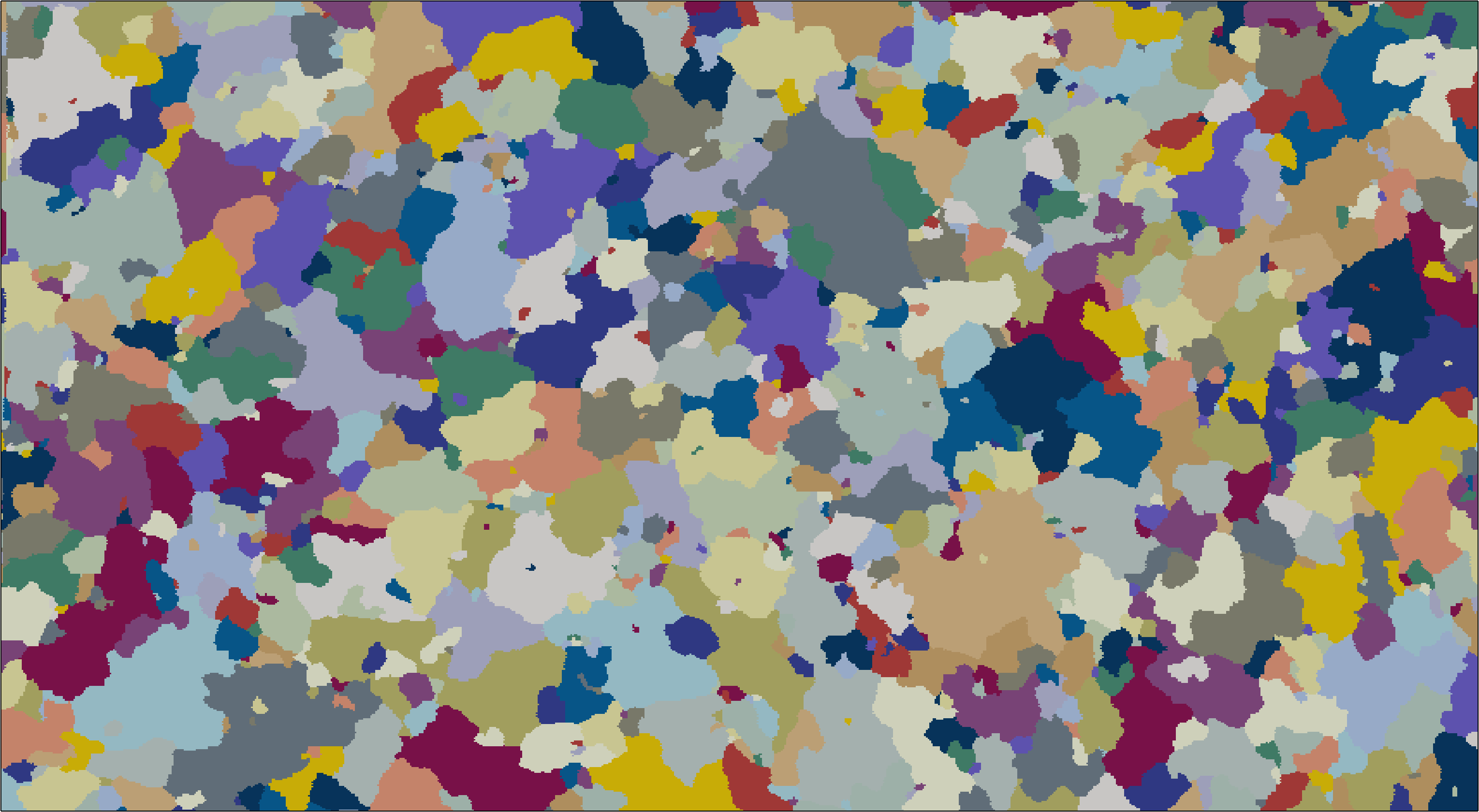};
\addplot[opacity=0.85] graphics [includegraphics cmd=\pgfimage,xmin=0, xmax=858, ymin=0, ymax=551.4] {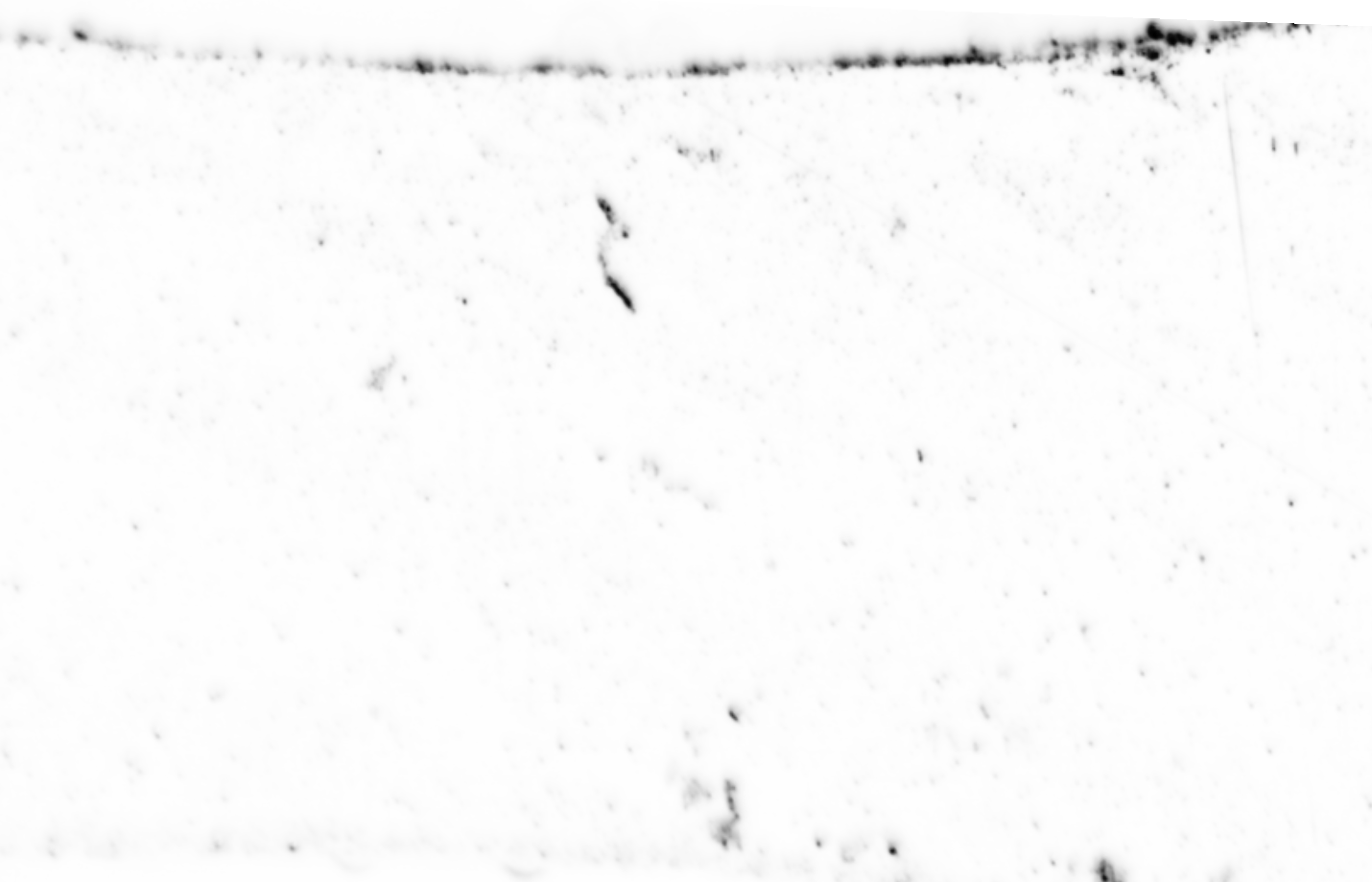};
\end{axis}

\end{tikzpicture}} \\ $\SI{8.00e6}{}$ cycles} &
        \specialcell{\hflip{
\begin{tikzpicture}

\begin{axis}[
axis equal image,
hide x axis,
hide y axis,
tick align=outside,
tick pos=left,
x grid style={white!69.0196078431373!black},
xmin=660, xmax=720,
xtick style={color=black},
y grid style={white!69.0196078431373!black},
ymin=118, ymax=200,
ytick style={color=black}
]
\addplot graphics [includegraphics cmd=\pgfimage,xmin=0, xmax=856.20003, ymin=30.484095, ymax=500.708367768879] {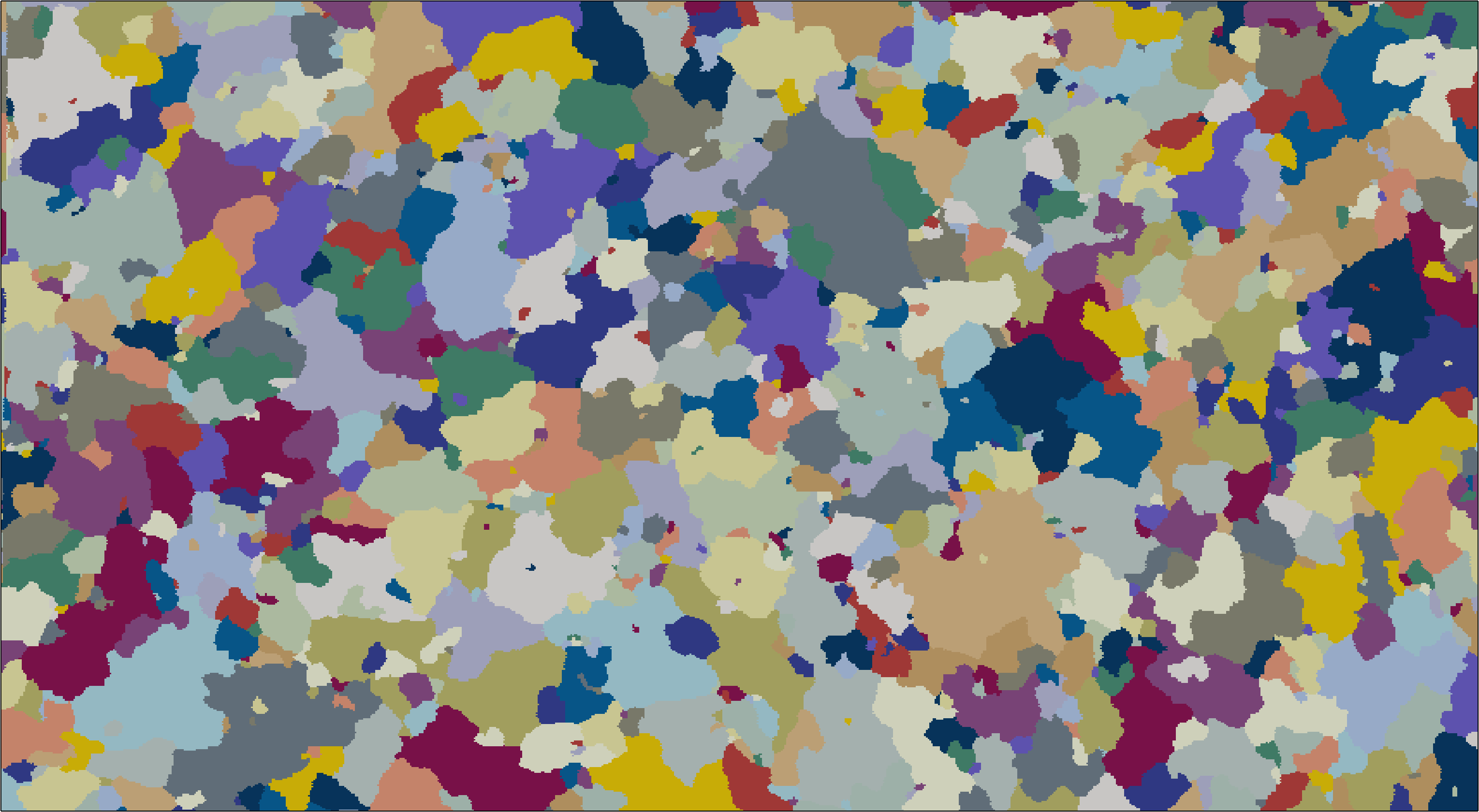};
\addplot[opacity=0.85] graphics [includegraphics cmd=\pgfimage,xmin=0, xmax=858, ymin=0, ymax=551.4] {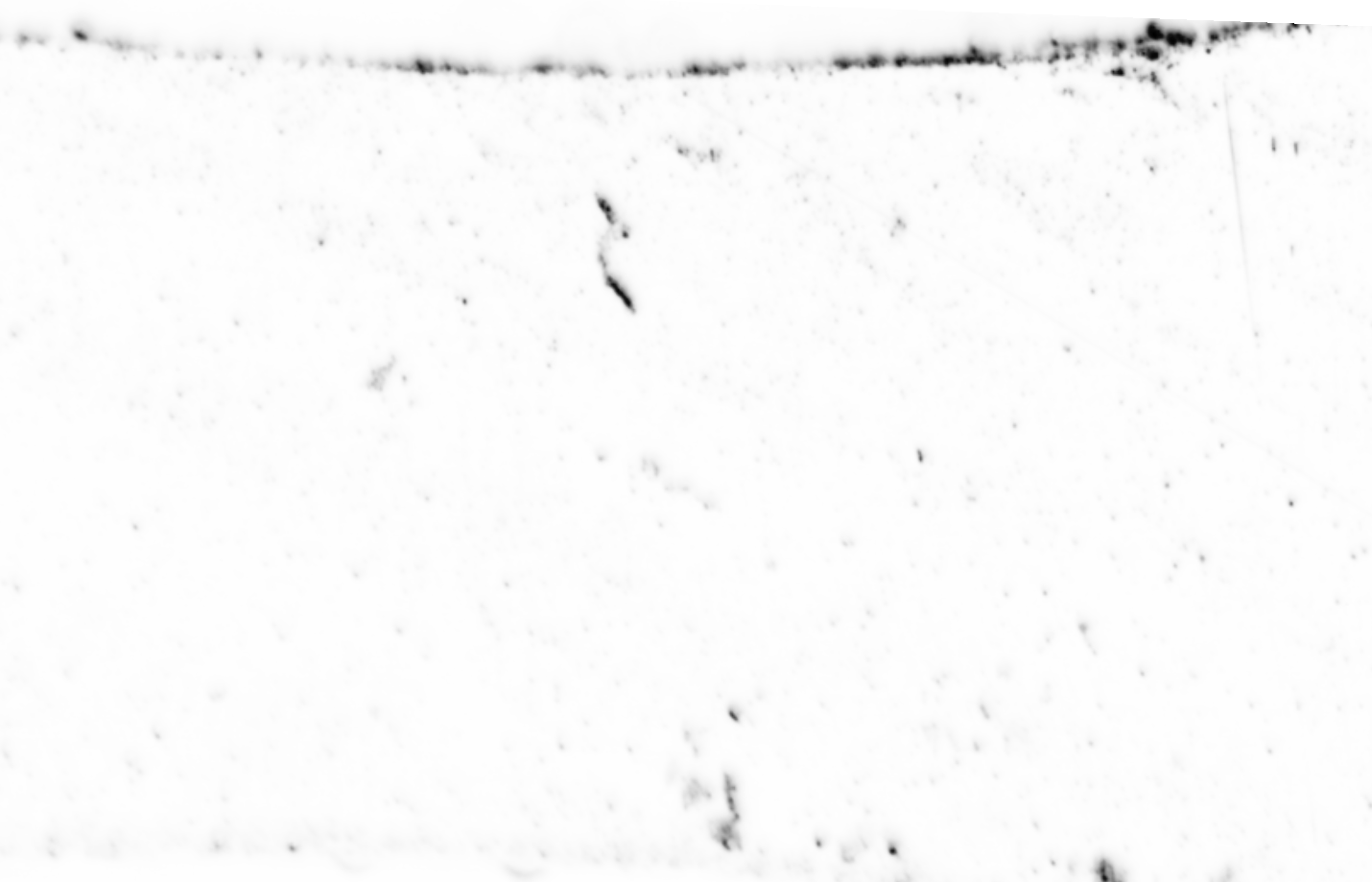};
\end{axis}

\end{tikzpicture}} \\ $\SI{8.98e6}{}$ cycles} &
        \specialcell{\hflip{
\begin{tikzpicture}

\begin{axis}[
axis equal image,
hide x axis,
hide y axis,
tick align=outside,
tick pos=left,
x grid style={white!69.0196078431373!black},
xmin=660, xmax=720,
xtick style={color=black},
y grid style={white!69.0196078431373!black},
ymin=118, ymax=200,
ytick style={color=black}
]
\addplot graphics [includegraphics cmd=\pgfimage,xmin=0, xmax=856.20003, ymin=30.484095, ymax=500.708367768879] {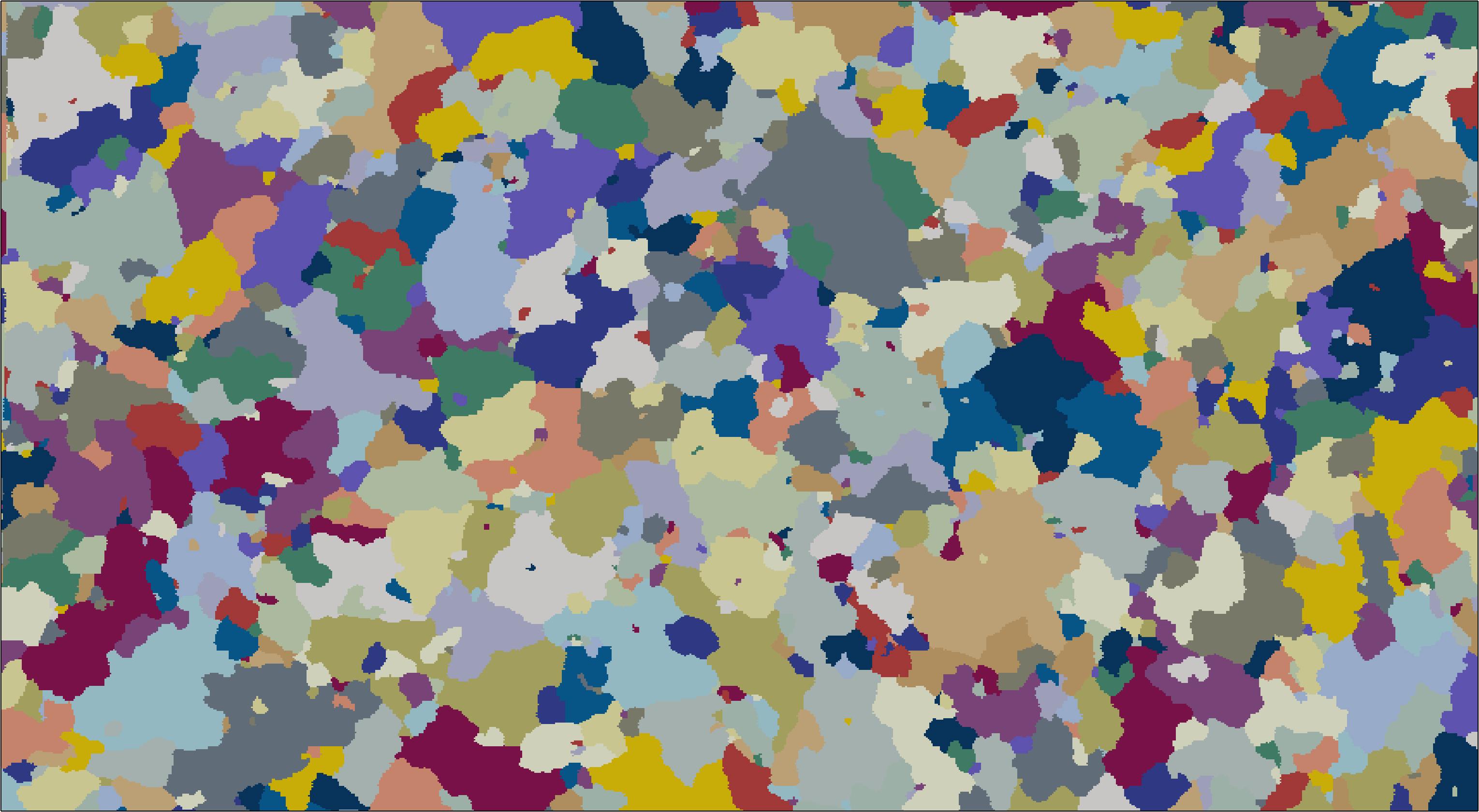};
\addplot[opacity=0.85] graphics [includegraphics cmd=\pgfimage,xmin=0, xmax=858, ymin=0, ymax=551.4] {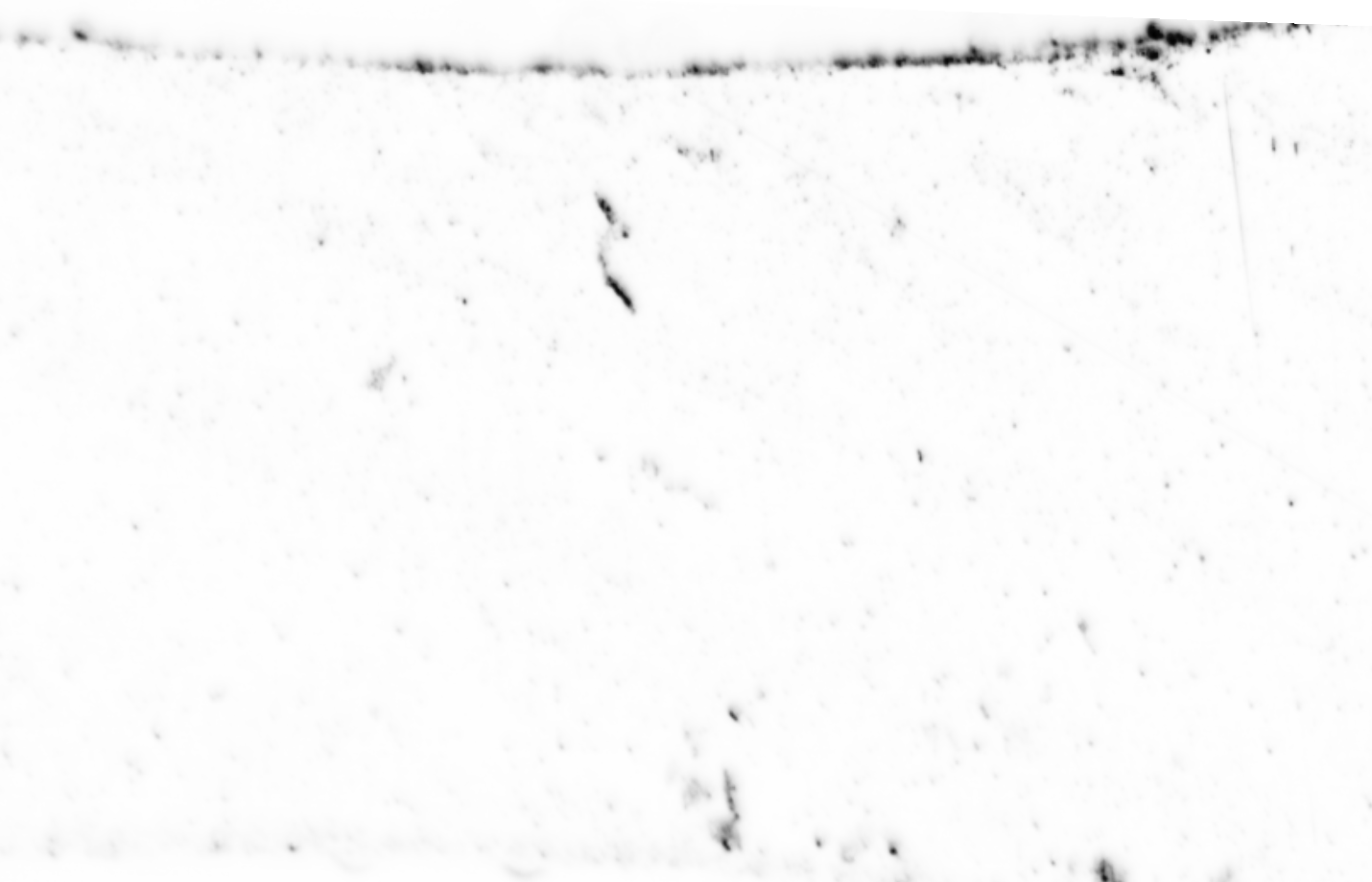};
\end{axis}

\end{tikzpicture}} \\ $\SI{1.01e7}{}$ cycles} &
        \specialcell{\hflip{
\begin{tikzpicture}

\begin{axis}[
axis equal image,
hide x axis,
hide y axis,
tick align=outside,
tick pos=left,
x grid style={white!69.0196078431373!black},
xmin=660, xmax=720,
xtick style={color=black},
y grid style={white!69.0196078431373!black},
ymin=118, ymax=200,
ytick style={color=black}
]
\addplot graphics [includegraphics cmd=\pgfimage,xmin=0, xmax=856.20003, ymin=30.484095, ymax=500.708367768879] {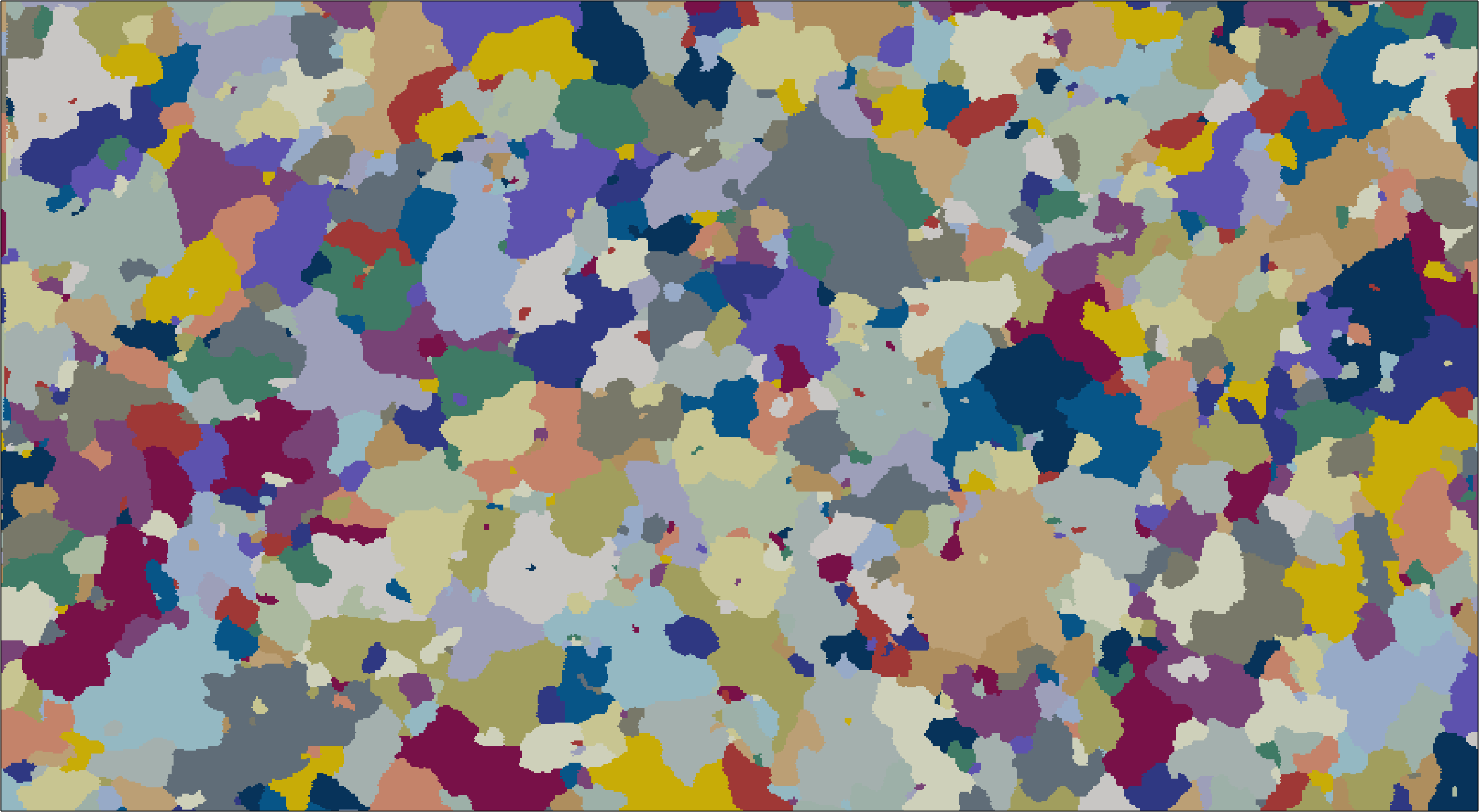};
\addplot[opacity=0.85] graphics [includegraphics cmd=\pgfimage,xmin=0, xmax=858, ymin=0, ymax=551.4] {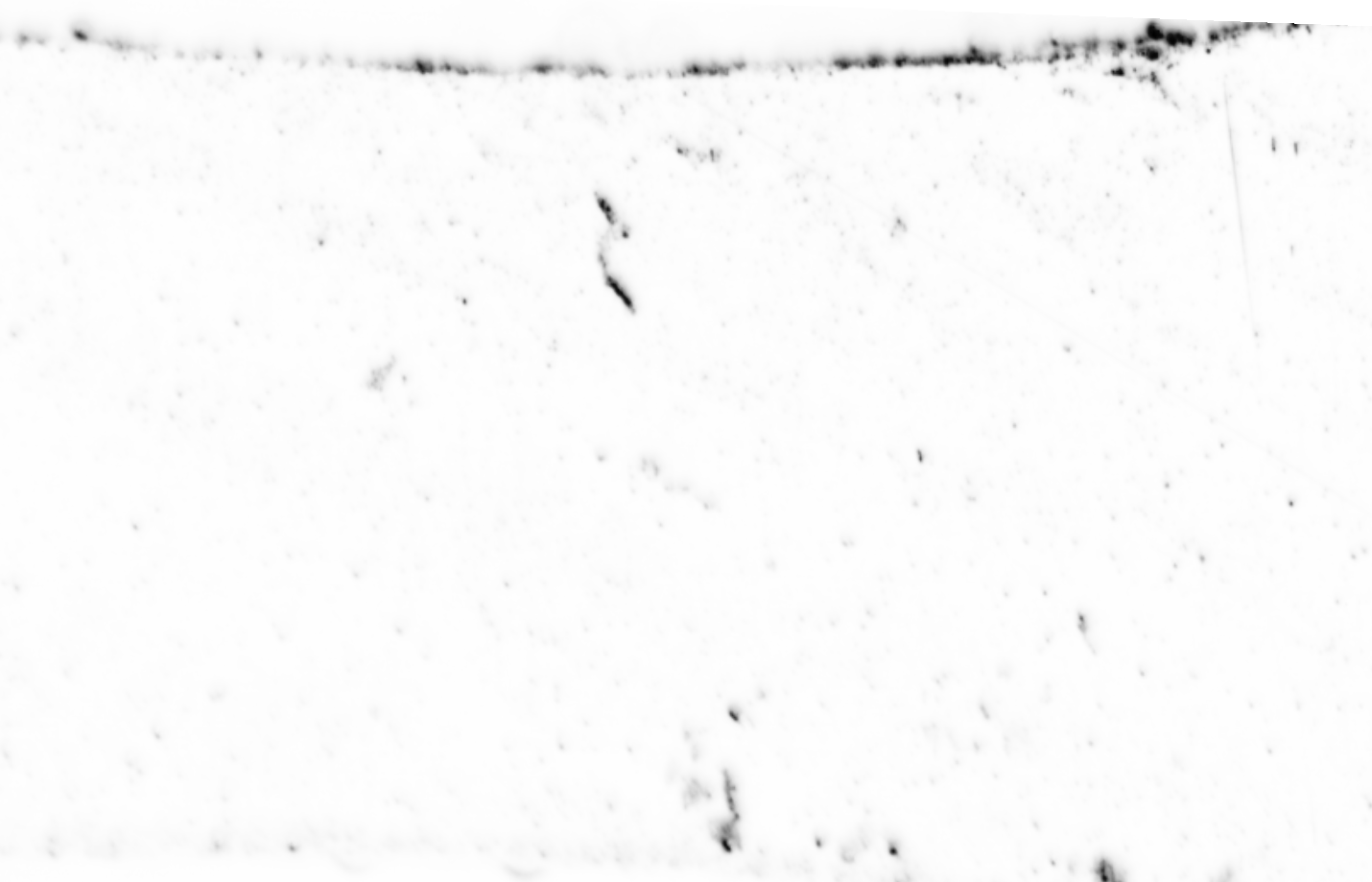};
\end{axis}

\end{tikzpicture}} \\ $\SI{1.15e7}{}$ cycles}  \\
        \specialcell{\hflip{
\begin{tikzpicture}

\begin{axis}[
axis equal image,
hide x axis,
hide y axis,
tick align=outside,
tick pos=left,
x grid style={white!69.0196078431373!black},
xmin=660, xmax=720,
xtick style={color=black},
y grid style={white!69.0196078431373!black},
ymin=118, ymax=200,
ytick style={color=black}
]
\addplot graphics [includegraphics cmd=\pgfimage,xmin=0, xmax=856.20003, ymin=30.484095, ymax=500.708367768879] {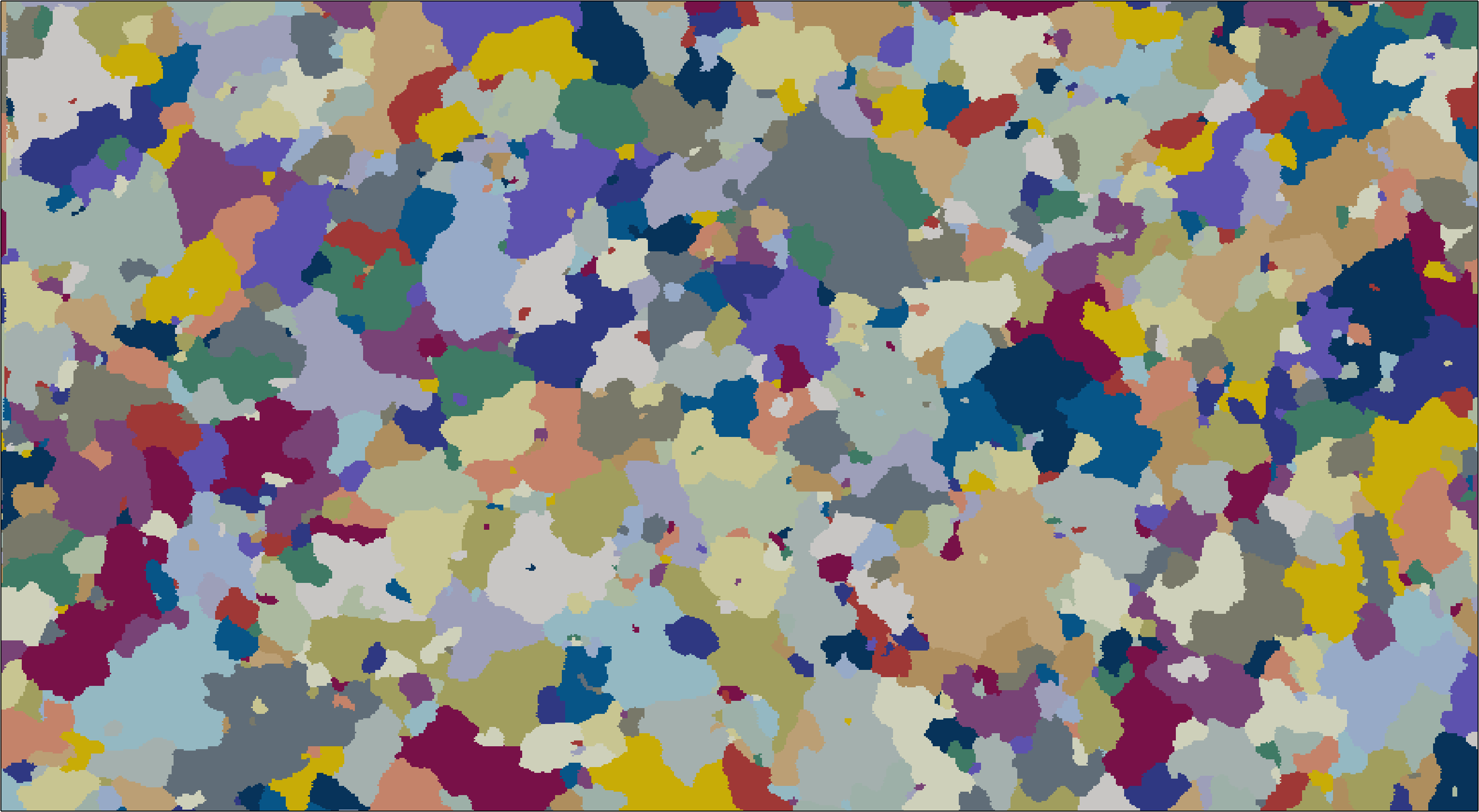};
\addplot[opacity=0.85] graphics [includegraphics cmd=\pgfimage,xmin=0, xmax=858, ymin=0, ymax=551.4] {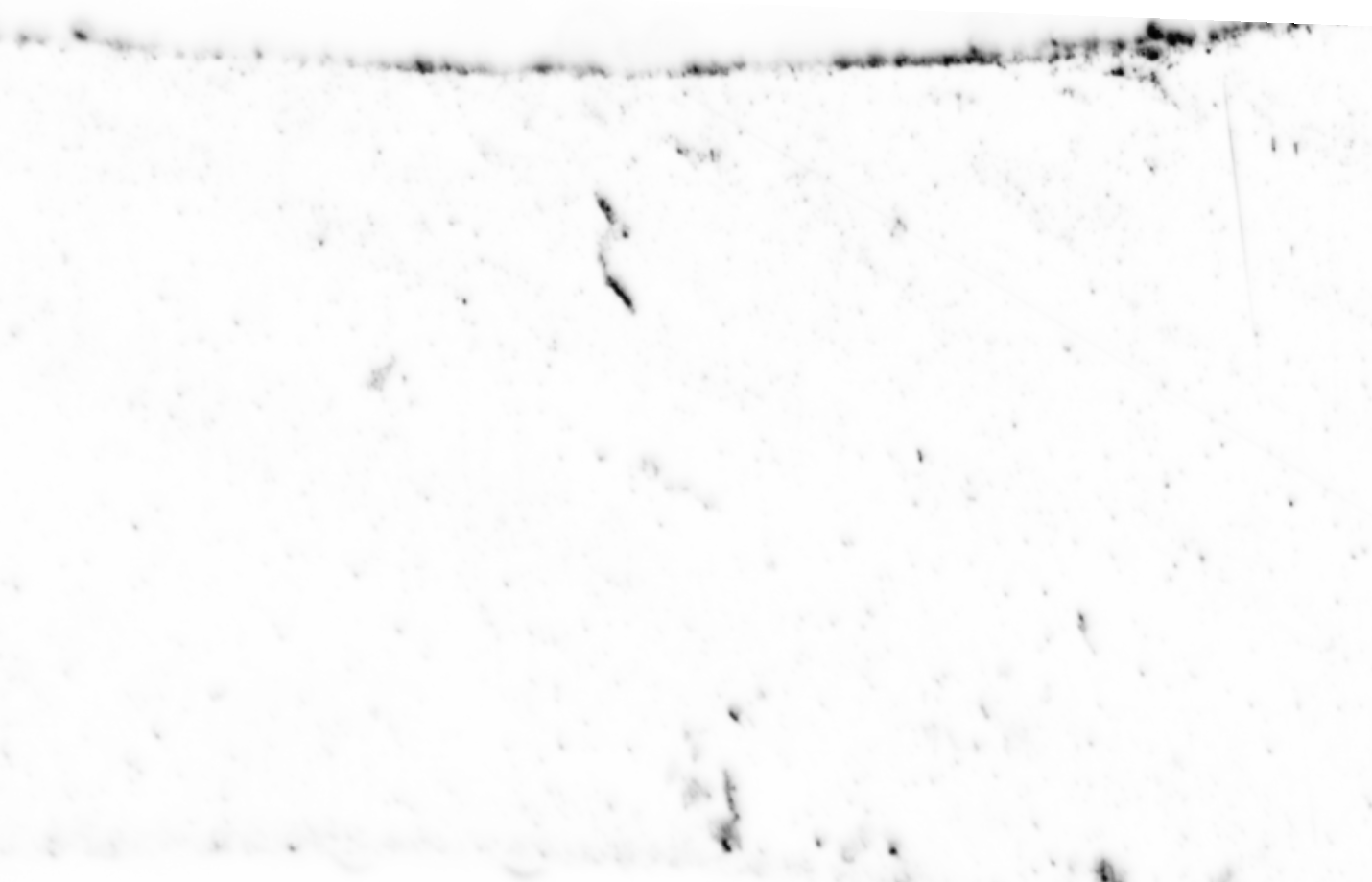};
\end{axis}

\end{tikzpicture}} \\ $\SI{1.21e7}{}$ cycles}  &
        \specialcell{\hflip{
\begin{tikzpicture}

\begin{axis}[
axis equal image,
hide x axis,
hide y axis,
tick align=outside,
tick pos=left,
x grid style={white!69.0196078431373!black},
xmin=660, xmax=720,
xtick style={color=black},
y grid style={white!69.0196078431373!black},
ymin=118, ymax=200,
ytick style={color=black}
]
\addplot graphics [includegraphics cmd=\pgfimage,xmin=0, xmax=856.20003, ymin=30.484095, ymax=500.708367768879] {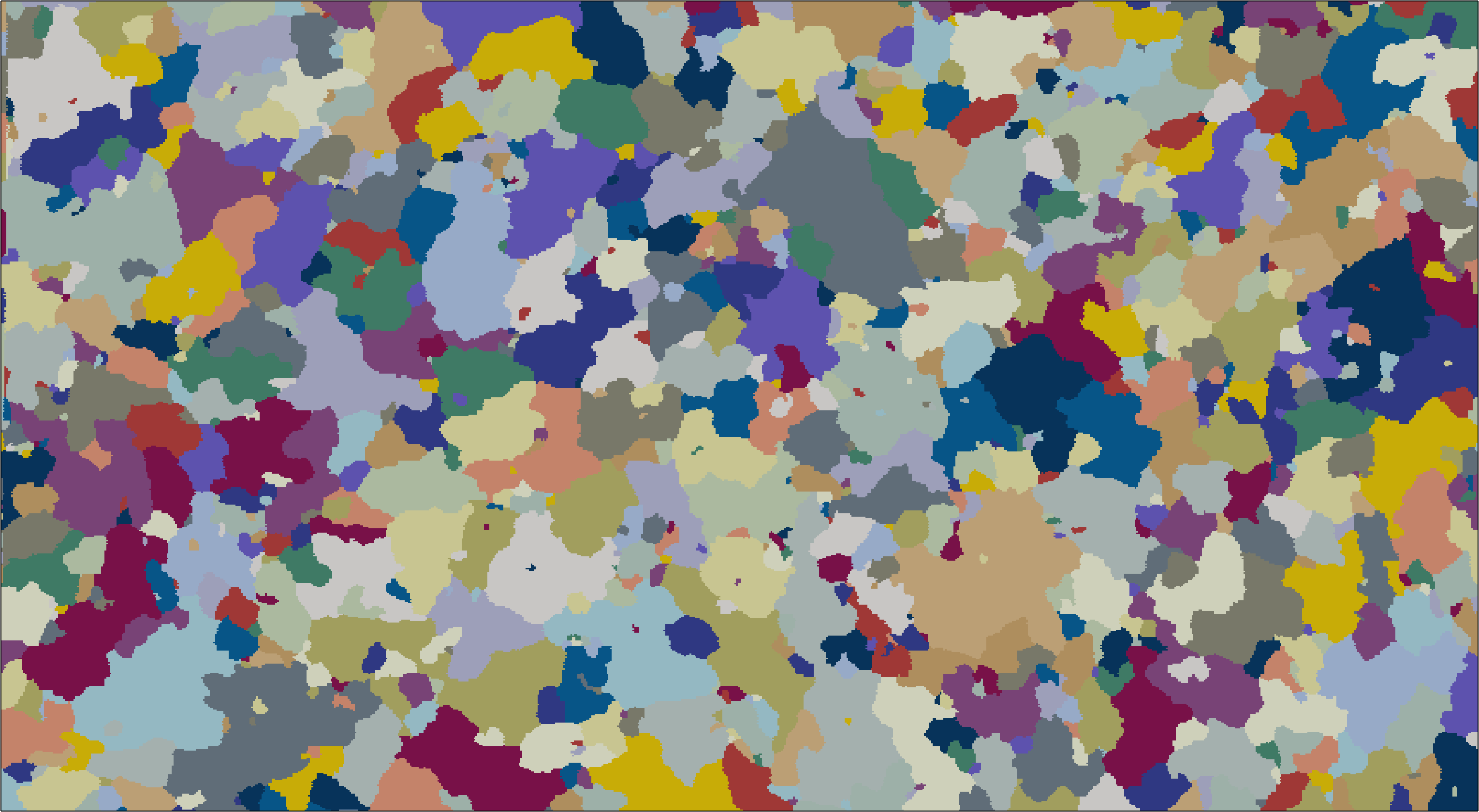};
\addplot[opacity=0.85] graphics [includegraphics cmd=\pgfimage,xmin=0, xmax=858, ymin=0, ymax=551.4] {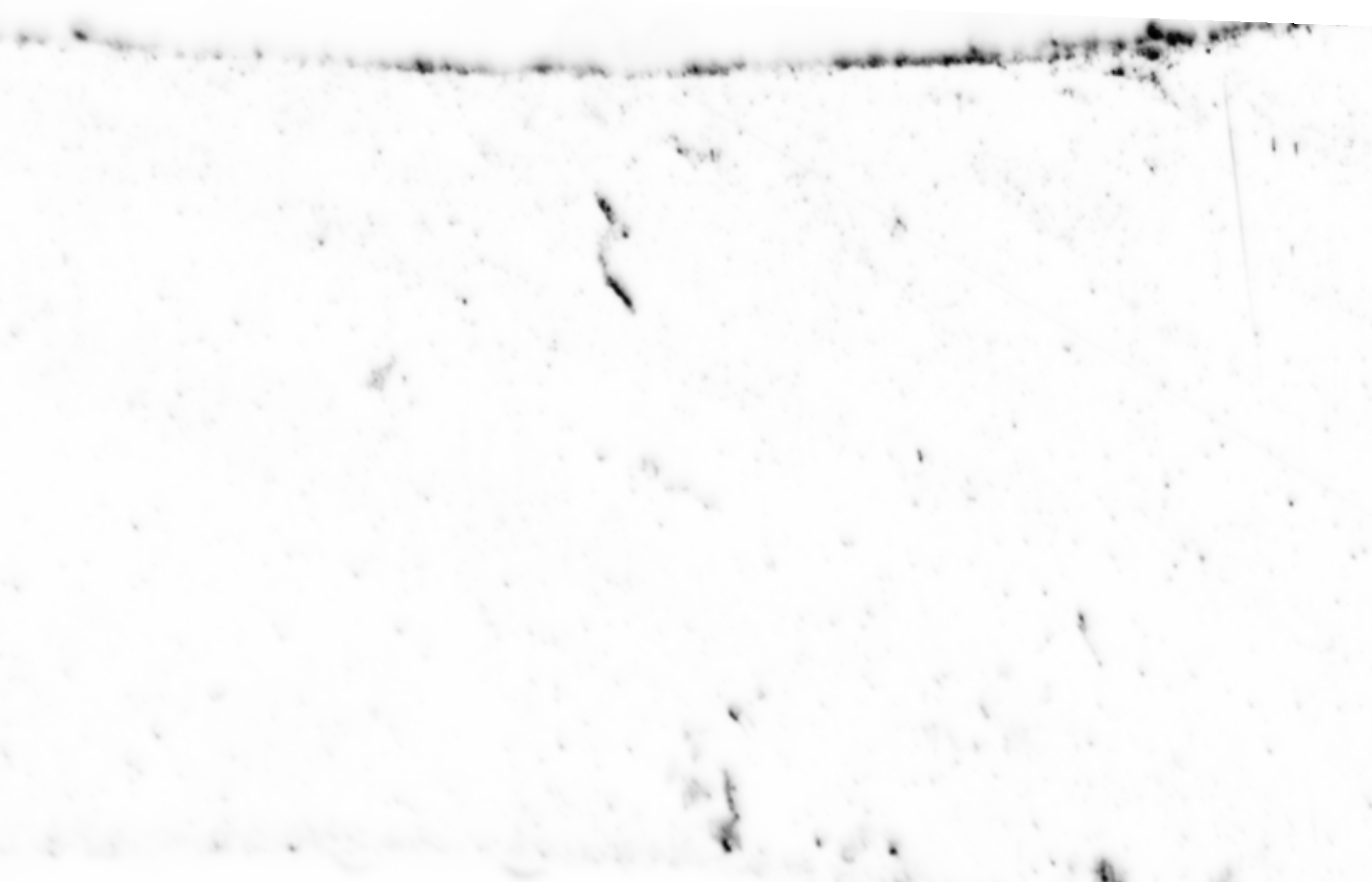};
\end{axis}

\end{tikzpicture}} \\ $\SI{1.42e7}{}$ cycles}  &
        \specialcell{\hflip{
\begin{tikzpicture}

\begin{axis}[
axis equal image,
hide x axis,
hide y axis,
tick align=outside,
tick pos=left,
x grid style={white!69.0196078431373!black},
xmin=660, xmax=720,
xtick style={color=black},
y grid style={white!69.0196078431373!black},
ymin=118, ymax=200,
ytick style={color=black}
]
\addplot graphics [includegraphics cmd=\pgfimage,xmin=0, xmax=856.20003, ymin=30.484095, ymax=500.708367768879] {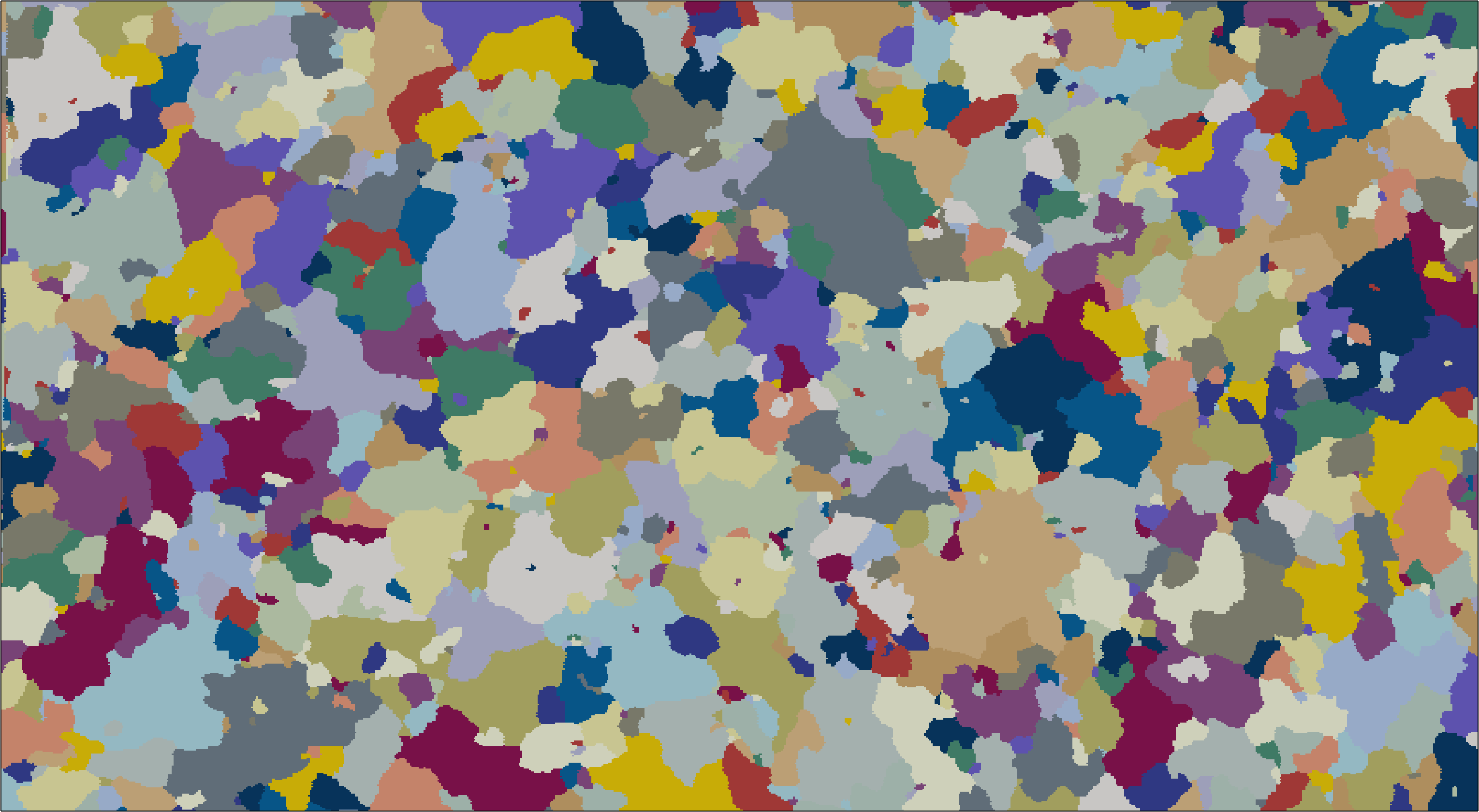};
\addplot[opacity=0.85] graphics [includegraphics cmd=\pgfimage,xmin=0, xmax=858, ymin=0, ymax=551.4] {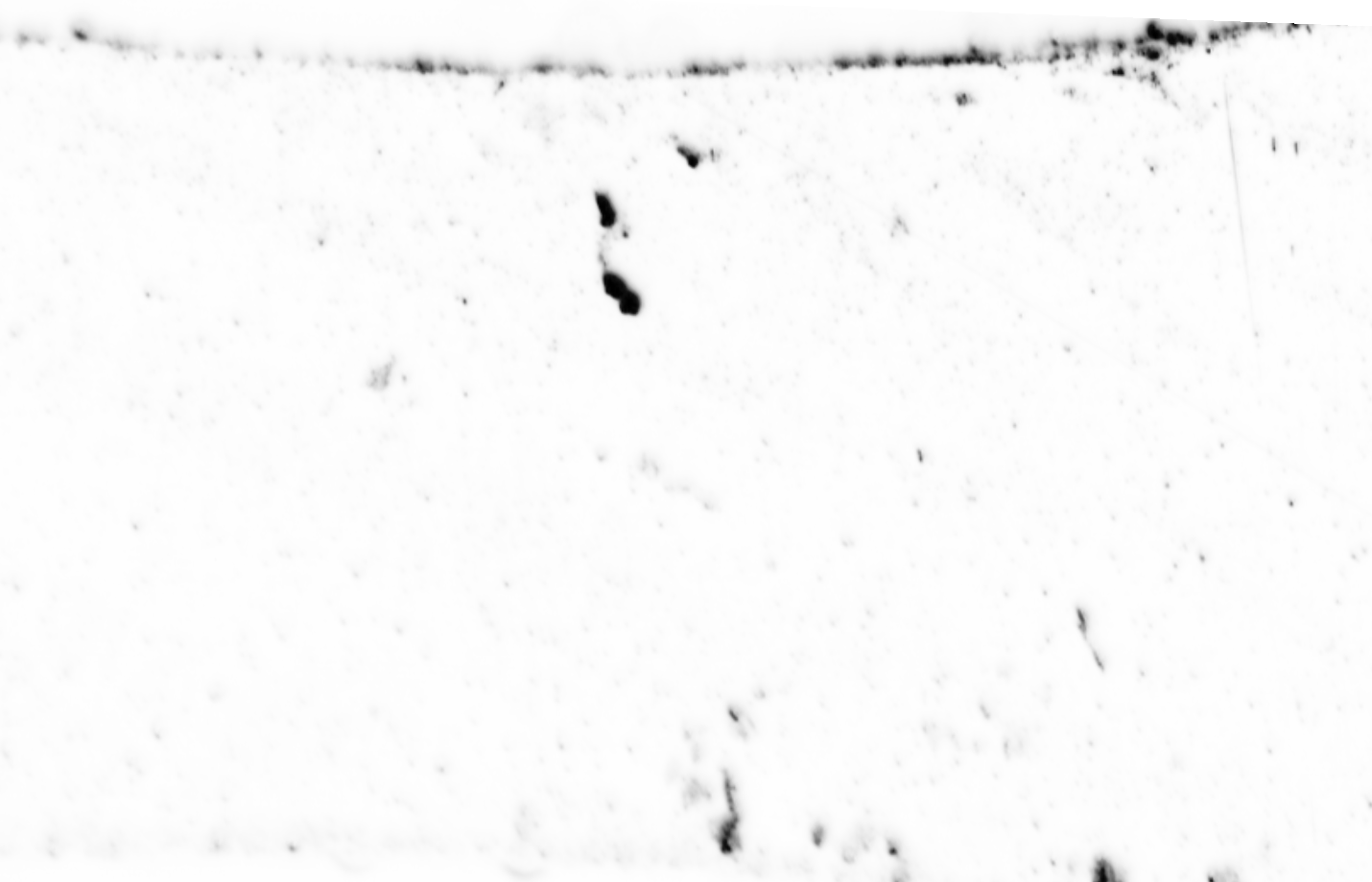};
\end{axis}

\end{tikzpicture}} \\ $\SI{5e8}{}$ cycles}  &
        \specialcell{\hflip{
\begin{tikzpicture}

\begin{axis}[
axis equal image,
hide x axis,
hide y axis,
tick align=outside,
tick pos=left,
x grid style={white!69.0196078431373!black},
xmin=660, xmax=720,
xtick style={color=black},
y grid style={white!69.0196078431373!black},
ymin=118, ymax=200,
ytick style={color=black}
]
\addplot graphics [includegraphics cmd=\pgfimage,xmin=0, xmax=856.20003, ymin=30.484095, ymax=500.401132046717] {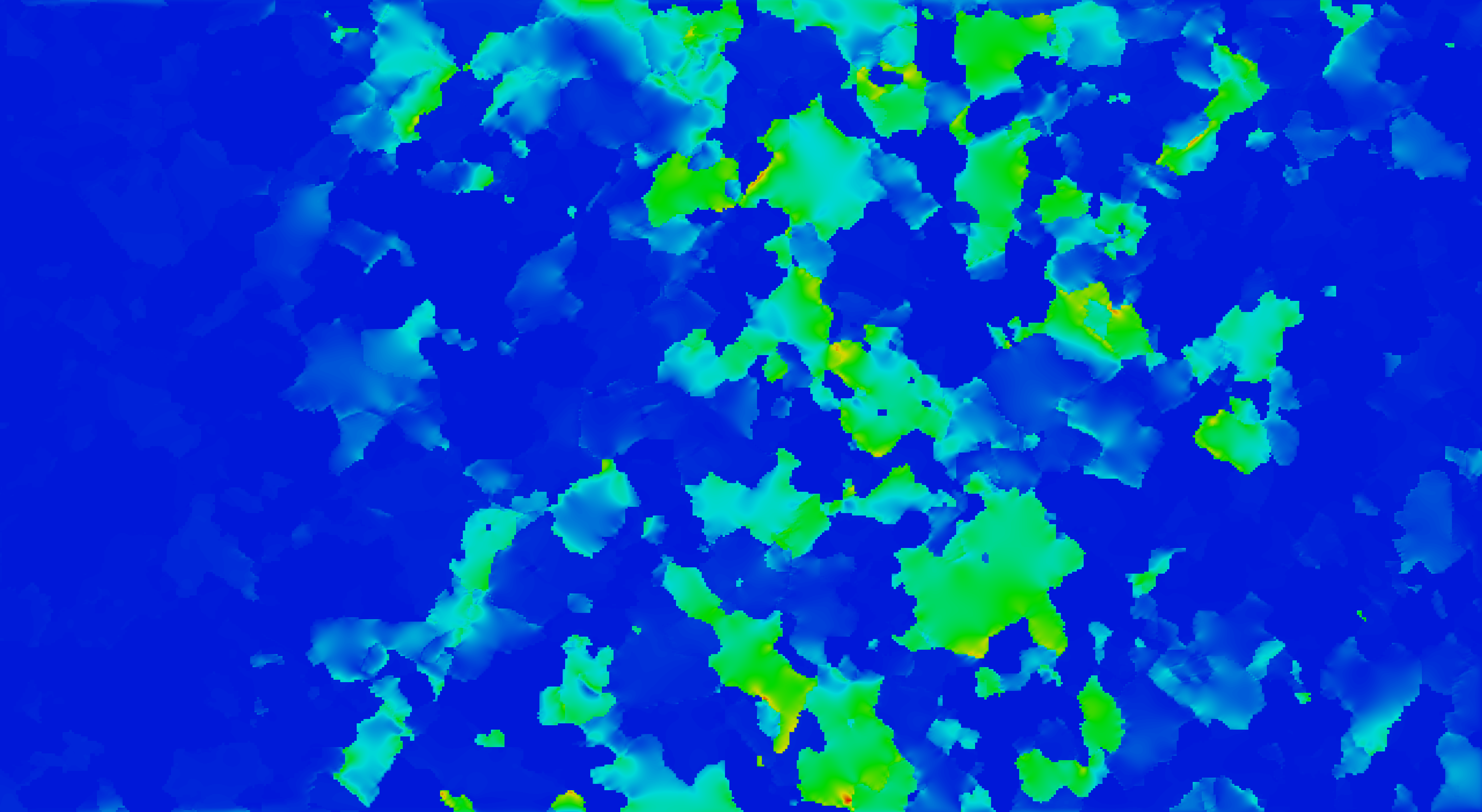};
\addplot[opacity=0.5] graphics [includegraphics cmd=\pgfimage,xmin=0, xmax=858, ymin=0, ymax=551.4] {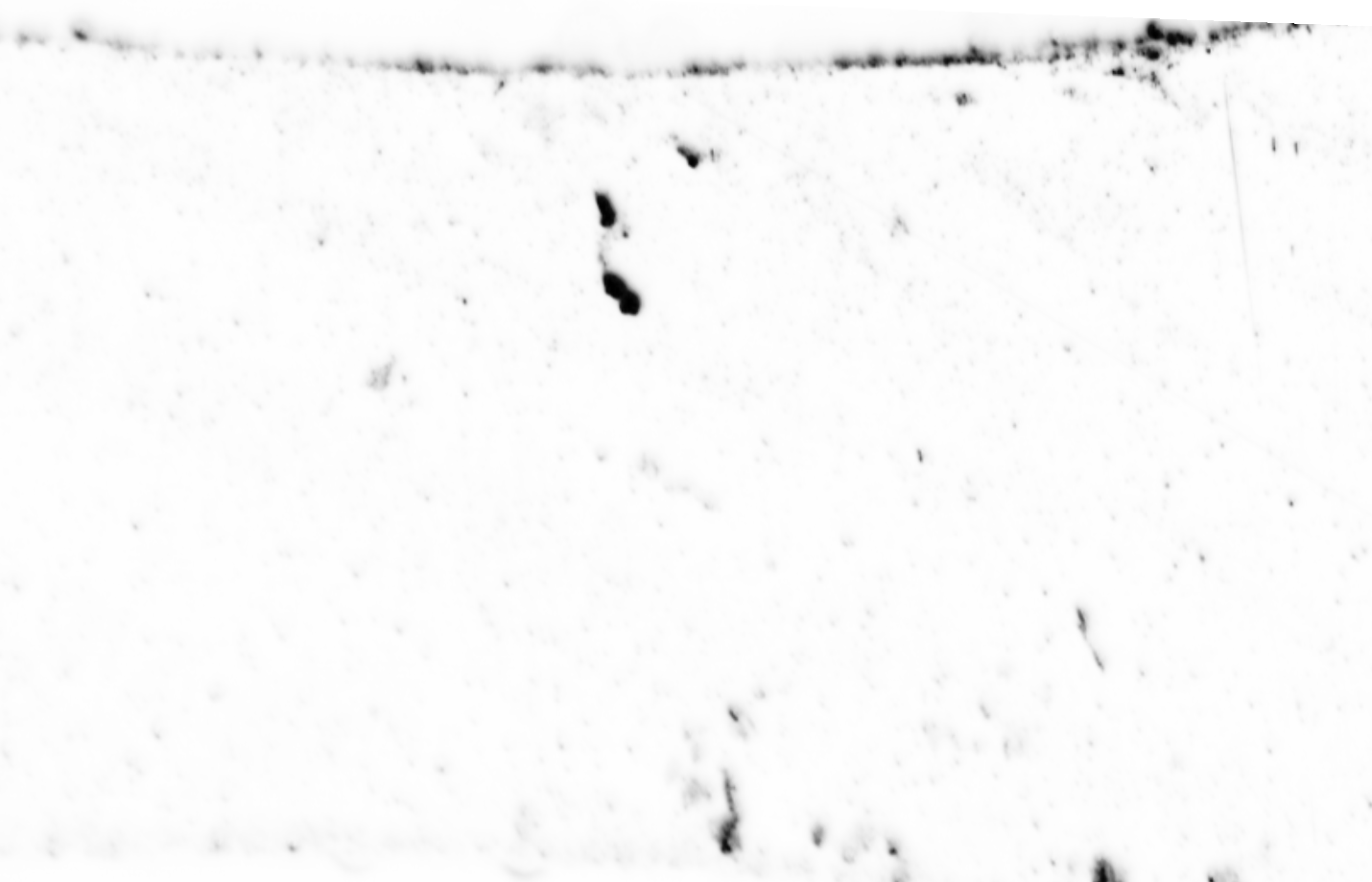};
\end{axis}

\end{tikzpicture}} \\ $\SI{5e8}{}$ cycles, $\fippover$}  \\
        \end{tabularx}
		\caption{}
	\end{subfigure}
	\hspace{2mm}
	\begin{subfigure}{0.325\textwidth}
		\def\svgwidth{1.0\textwidth}
		\footnotesize
		\import{figures/resultsnew/}{Time_series_analysis_2_render.tex}
		\caption{}
	\end{subfigure}
	\caption{a) Light optical grayscale image series superimposed onto the region annotated with the black box and arrow in Figure \ref{fig:SDV115_Overlayed} to track crack formation and evolution. The underlying image represents the meshed microstructure in grain color coding and for the last frame, the $\fippover$-heatmap is shown in addition. The numbers indicate the corresponding load cycle numbers. b) Secondary electron SEM image of the final crack state (1.1$\times{}10^9$ cycles). Numbering and arrows illustrates the crack initiation and growth process. Grain boundaries analogous to a) are faintly visible through discontinuities in the grayscale contrast.}
	\label{fig:time_series}
\end{figure}

\section{Discussion} \label{discussion}
The primary objective of developing a validation methodology for fatigue damage prediction of micromechanical simulation models was satisfied. At its core, the validation evolves around automatized localization of damage locations with a convolutional neural network (1) described in \cite{Thomas2020}, multimodal image registration techniques (2), and a CPFE sub-modeling approach (3). Each of these three core elements, at their current stage, introduces specific beneficial aspects as well as limitations to the validation approach.

\subsection{Validation methodology}

From the comparison of Figure \ref{fig:time_series} a) and b) it is evident that secondary electron SEM images can provide comparatively more detail and an appropriate damage type distinction as opposed to the light optical image modality. For instance, such SEM images permit distinguishing the crack from the plasticity traces surrounding it. Since topography-information is recovered from the Everhart-Thornley SEM and applied light optical imaging techniques, only surface-altering damage exceeding these techniques' respective detection limit can be registered. A convolutional neural network based on a U-Net architecture \cite{Ronneberger2015} was trained to detect damage locations of the classes protrusion and cracks based on such topography sensitive SEM images \cite{Thomas2020}. While this model can detect such micromechanical damage reliably, individual damage object boundaries are difficult to segment accurately due to the characteristics of protrusions, and the imaging methodology \cite{Thomas2020}. To counteract this, a visual inspection and minor adjustments of the U-Net segmentation predictions were performed for the ground truth. Additionally, damage masks derived from the overlayed light optical time series data in Figure \ref{fig:time_series} a) through thresholding were considered (not shown). In that case, the SEM-derived semantic segmentation mask was used to select optimal thresholds for light optical difference images by maximizing the Jaccard index \cite{jaccard1902lois} between the SEM-derived mask and the one of the last light optical image. Therefore, a damage category-dependent threshold was selected and applied throughout the whole time series to assimilate the damage masks derived from both modalities. While their derived masks show similar patterns, damage instances appear more extensive overall in the masks derived from the light optical time series. Deviations between both damage masks exist particularly at the specimen edge. The accordance of both maps is decent but the SEM-derived map, even prior to expert corrections, provides a higher fidelity. Since damage locations in many materials tend to adhere to grain boundaries, a reliable damage detection methodology is a fundamental requirement to enable an appropriate assignment to microstructural features.

The same applies to the registration of the heterogeneous data, where the applied methodology was found to yield satisfying relative alignment between damage locations and microstructure, see \cite{Durmaz2021}. In materials with characteristic grain sizes below the micron scale, where damage is comparatively more confined, reliable damage detection proved to be more difficult, see \cite{Thomas2020} under the conditions outlined there. Moreover, such materials pose comparatively higher requirements on the data registration methodology. Therefore, the investigation of small grain materials presumably poses a limitation at the current stage. This, however, is not of concern here since the investigated material has an average grain size of 25\,$\upmu$m.

The sub-modeling approach enables translating various boundary conditions onto the microstructure domain to which the CPFE model is applied. While in this experimental case of VHCF-regime loading, the assumption of a linear elastic model on the mesoscale specimen level appears appropriate and the majority of the plasticity occurs within the microstructurally modelled CPFE domain, larger deformations could necessitate the transition to elastic-plastic models on the specimen scale and different registration methodologies. Since this model utilizes unidirectional coupling from the specimen to the microstructure scale, and not vice versa, microstructural short crack growth poses an upper bound, where the specimen-scale stress state can still be assumed to be maintained. In contrast to macroscopic specimens, the specimen-scale stress state can be altered at a relatively small absolute crack length due to the specimen's small dimensions. Remeshing-based approaches (\cite{Proudhon2016}) and XFEM techniques (\cite{Wilson2018}) can address this issue and further inherently model the stress relaxation in the vicinity of the crack.       

\subsection{Assessment of FIPs with respect to experimental damage locations}

While the accumulated plastic strain $\fipp$ is based on the notion that all slip systems contribute to fatigue crack nucleation, the Fatemi-Socie $\fipfs$ only takes the slip system with the maximum shear strain range into account. Hence, a single slip criterion is established in $\fipfs$ combining the effect of crystallographic slip and tensile stress on a slip plane. The measurement of dissipated energy in terms of $\fipw$ is similar to the $\fipp$, an accumulation of dissipated energy over all slip systems.

\begin{figure}[htbp]
	\centering
	\footnotesize
	\includegraphics[width=0.5\textwidth]{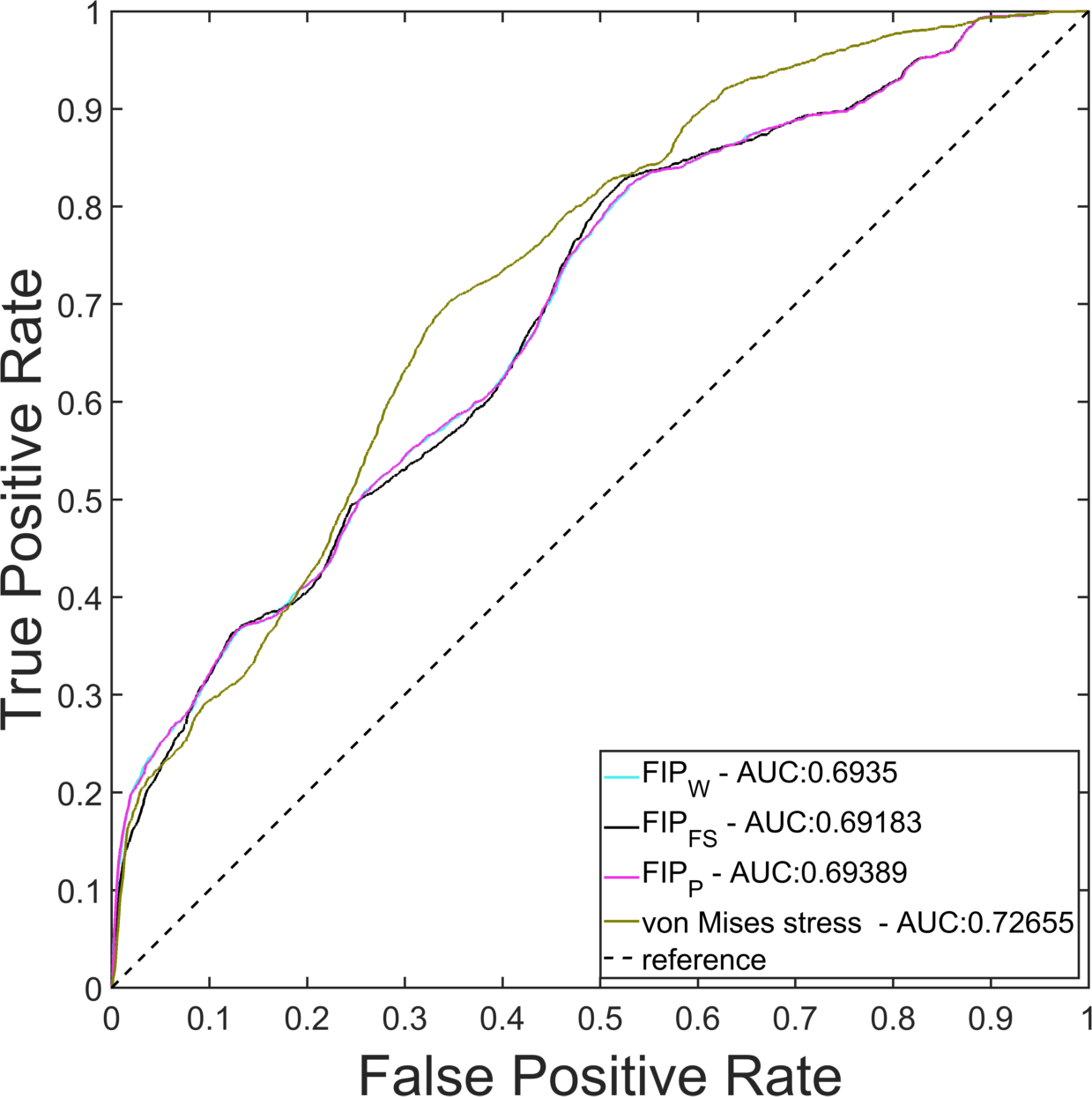} 
	\caption{Receiver operating characteristic (ROC) curve of normalized FIP metrics and normalized von Mises stress using the segmented damage map as ground truth. This curve is constructed by sweeping the decision boundary from zero to unity and evaluating the false positive and true positive rate for each data point. The legend additionally provides the area under the ROC curve (AUC) which is a metric for the binary classification accuracy. The reference diagonal represents a classifier with no predictive power, an AUC of unity would correspond to ideal distinction of damage and no damage by the metric. Note that $\fipw$ and $\fipp$ almost coincide.}
	\label{fig:FIP_violin}
\end{figure}

To achieve a quantitative comparison of the damaged area between experiment and simulation, a threshold must be applied to the FIP fields. The selection of an appropriate FIP metric threshold is hampered by the long-tailed FIP distribution arising due to the VHCF loading and the large number of grains. The selection of the most suitable metric and its ideal threshold can be facilitated by the receiver operator characteristic illustrated in Figure \ref{fig:FIP_violin}. The curve is computed by thresholding the normalized FIP metrics and von Mises stress at different values and assessing the confusion matrix elements (true-positives, true-negatives, false-positives, and false-negatives) to infer the true-positive and false-positive rates at each threshold. Therefore, the normalized damage metrics are considered class probabilities, where the boundaries zero and unity represent undamaged and damaged pixels, respectively. The SEM-derived segmentation map is taken as ground truth for the construction of this diagram, and cracks, as well as protrusions, are considered collectively as damage to formulate a binary classification problem. The area under the ROC curves is a metric (ROC-AUC) for the predictive power of the CPFEM model and the FIP formulations at hand for the given material. The same procedure is followed for the microstructure-sensitive von Mises stress. In the ROC curve, the $\fipw$ and $\fipp$ are nearly congruent, which is comprehensible, taking into account the similarity of their formulation. The ROC curve for $\fipfs$ is discernable from the former two but yields similar ROC-AUC values of approximately 0.69. Interestingly, the micromechanical von Mises stress has more predictive power achieving a ROC-AUC of almost 0.73.    

The surprisingly good prediction accuracy of the von Mises stress relative to the FIPs might be explained by differing mechanisms in the LCF domain and the HCF/VHCF regimes. In the Coffin-Manson-Basquin equation, the latter are governed by the elastic strain amplitude received from Basquin's law and thus dominated by a term proportional to the stress amplitude, see \cite{radaj_ermudungsfestigkeit_2007}. Furthermore, introducing a term related to stress can also be seen as a motivation for the Fatemi-Socie and dissipated energy FIPs instead of accumulated plastic strain. \cite{Schafer2019} also noted the generally limited prediction accuracy of the FIPs in the HCF regime and better accuracy of $\fipfs$ and $\fipw$ as opposed to $\fipp$. However, their results also demonstrate the good lifetime prediction of the FIPs in the LCF domain.

In the FIP distributions, spatial discontinuities occur due to orientation differences between grains, i.e., at grain boundaries. Furthermore, the FIPs show a strong localization within grains towards GBs. While the FIPs reproduce the experimental tendency that damage adheres to GBs \cite{Durmaz2021}, they do not seem to most prominently indicate the correct GB locations.  

Moreover, the damage segmentation map contains few damage locations at regions where each FIP predicts low values. Among this subset of damage locations, only the one shown in Figure \ref{fig:time_series} can be unambiguously ascribed to fabrication-induced pore defects, which were not considered in the modeling. Analogously, it cannot be excluded that elongated MnS inclusions, which are known to exist in this material, could be present beneath the specimen surface. While such defects could clearly affect damage emergence, their spatial distribution in the microstructure renders it unlikely that such defects contribute to a large portion of observed damage locations. Additional nanoscopic computer tomography experiments or similar would be required to capture those defects. Similarly, the OPS treatment results in a polishing relief, where surface steps of few tens of nm at grain boundaries (non-modelled) could contribute to damage emergence.

This leads to the conclusion that the assumptions made in the CPFE model and FIP formulations do not represent microstructural fatigue damage emergence appropriately. There is a range of incompatibilities between modeling and experiment:

\begin{enumerate}
\item The simulation considers an ideal rectangle cross-section, while the specimen possesses rounded-off edges. This influences the local stress state. However, many damage locations are not situated at the specimen edge and thus only affected to a minor degree by this stress state discrepancy.  
\item Crack growth modeling is omitted. Therefore, stress release in the vicinity can not be captured by the simulation model.
\item  Grain morphology and crystallographic orientation of adjacent grains may cause elastic anisotropy as well as length scale-dependent phenomena. The used phenomenological CP model captures the elastic anisotropy. However, for the grain size effects such as the Hall-Petch relation, a non-local crystal plasticity formulation would be required.
\item Prismatic grains are assumed; however, there are many instances in the literature indicating the importance of GB inclination, e.g., \cite{Saylor2003}. Using a phenomenological model for FCC copper \cite{zeghadi2007ensemble_1, zeghadi2007ensemble_2} found that by constraining the surface microstructure and altering the sub-surface microstructure, the predicted surface plastic slip is strongly affected for uniaxial tension loading. Namely, fluctuations of more than 60\% were found for 40\% of the free surface area between different statistical volume elements with constrained surface microstructure. Moreover, in that work, the region beneath the surface that influences the surface plasticity field was twice the average grain size in depth. Despite not considering the change of the local grain boundary slip transmission characteristics when changing its inclination angle (see \cite{sangid2011a, sangid2011b}), similar to our study, they highlight the severe impact of the 3D orientation distribution. In contrast to their computational study, in our case, the influence of the 3D microstructure is somewhat alleviated since our specimen is driven in \textit{bending} resonant loading accentuating the stresses at the specimen surface. Moreover, recent computational studies suggest that the free surface's stress concentration is comparatively more pronounced when considering high symmetry crystal structures \cite{Stopka2020}, such as the BCC material characterized here.
\item Apart from the lack of 3D microstructure data, the employed Ohno-Wang kinematic hardening rule of the phenomenological CP model does not attempt to capture the influence of local slip transmission characteristics of grain boundaries on the resulting back stresses. However, these characteristics, amongst others, are presumably of high importance when it comes to localization of fatigue damage. During calibration of the Ohno-Wang kinematic hardening rule, the hardening parameters $\khone$, $\khtwo$ were tuned to fit the experimental macroscopic $\upsigma$-$\upepsilon$-Hysteresis in the LCF regime. There are several sets of $\khone$ and $\khtwo$, resulting in an adequate match. Nevertheless, only one set was taken into account in this study and used as the hardening law parameters. Possible deviations in microstructure resolved strains between these sets are neglected. An alternate calibration concept of the hardening model would be to utilize HR-DIC to directly capture local sub-grain strain fields, as shown in \cite{Zhang2016}. In \cite{Cerrone2015} a hardening rule was suggested that included a lattice incompatibility term adapted from \cite{Beaudoin2000} and a modified Voce-Kocks term, see \cite{Kocks1976}. In conjunction with similar FIPs, this hardening law yields hot spots following experimental damage and localizing to GBs in their small sample size study. However, it is unclear if these results also transfer to steel and a different parameter calibration procedure than the applied one is required for the application of this formulation.
\item Another possible cause for experimental damage not being reproduced by the FIPs is that they map the mechanical fields onto the tendency of protrusion/crack formation inaccurately. However, since the microscopic deformation field has not been validated, it is debatable whether the underlying CP models, the FIP formulations or both lead to mediocre localization. In \cite{Chen2018a}, amongst others, the same FIPs as in this work were investigated. This work can confirm their observation of corresponding FIP hot spots showing a discrepancy to experimental damage sites. Furthermore, \cite{Chen2018a} proposed an alternative FIP that succeeded in predicting fatigue crack localization in the vicinity of inclusions. This FIP is based on the stored energy per cycle. However, for this FIP computation, a non-local model capable of estimating the geometrically necessary dislocation density is required. The application of such a formulation would severely increase computational cost.   
\item The hardening model parameters are derived from a low strain rate experiment, but the target fatigue experiment was cycled at frequencies of $\approx$ 2\,kHz. Therefore, the question arises whether the calibration results in a phenomenological hardening model that adequately represents the material's dislocation dynamics. It is known that through the characteristic thermally-activated screw dislocation glide, strain rate and temperature effects play an essential role in the plasticity and fatigue of BCC materials \cite{Cereceda2016, Geilen2020}. Furthermore, the strain rate sensitivity exponent is chosen in a way that strain rate effects are neglected.
Typically specimen heating affects high-frequency fatigue testing severely since frictional losses constitute approximately 95\% of the dissipated energy during plastic deformation \cite{Kamlah1997, Hodowany2000}. In contrast, for our testing methodology, we anticipate that the pronounced surface-to-volume ratio of the mesoscale specimen, the low strain amplitudes, and forced convection through specimen motion permit cycling at faster rates without notable heating on the global specimen scale. Nonetheless, thermal hotspots that alter the dislocation dynamics locally can not be excluded.
\item Even for low strain rates and room temperature, additional activation of $\{112\}$ slip systems for $\upalpha$-iron in monotonic tensile and compression tests was previously reported by \cite{Franciosi2015}. Therefore, the applied restriction to the $\{110\}$ family of slip planes in modeling could be called into question. The overall neglect of the non-Schmid behavior, see \cite{Franciosi2015, Cereceda2016} and cross-slip, a typical mechanism of screw dislocation movement in BCC materials, are potentially responsible for poor localization.
\item Additionally, the initial dislocation density, which may be of importance, is not respected. While the EBSD data step size is selected such that a geometrically necessary dislocation density estimation can be performed \cite{pantleon2008resolving, jiang2013measurement}, the statistically stored dislocation density can not be assessed with the dataset provided here. With further parametrization efforts, a non-local crystal plasticity model could take dislocation densities into consideration. However, such a model, for the present model size, might not be feasible from a computational cost point of view.
\end{enumerate}


In conclusion, it can be stated that there are multiple modeling choices, which can result in such a discrepancy to the experimental observations. These require systematic exclusion by conducting intermediate validation of strain fields, including further relevant microstructural information and potentially adapting the hardening and flow rules as well as FIP formulations. Thereby, a suitable compromise between model fidelity and computational as well as experimental model parametrization effort can be achieved.

The multimodal experimental data set stores the reality for future validations, allows for sensitivity studies to incorporate relevant features into the model purposefully, and enables, e.g., Bayesian optimization of model parameters to the experimental observation. This ultimately can lead to an improved discriminability of damaged and non-damaged regions in future predictions. The validation concept composed of deep learning damage segmentation, multimodal data registration, and submodeling is assumed to be applicable to a wide range of materials. Arguably, the deep learning networks trained in a supervised fashion pose the limiting factor in this regard since their generalization capability across datasets is known to be compromised \cite{Thomas2020}. However, novel machine learning concepts such as unsupervised domain adaptation \cite{goetz2021addressing} can render such models transferable to a wider range of materials and damage types without requiring annotated data for new target materials.     

All necessary data and code to carry out micromechanical simulations and compare the results to the acquired experimental data is open sourced in \cite{fordatis_repo} and \cite{github_repo}. The provided material includes the damage time series data, EBSD data, the meshed macro model with boundary conditions and the meshed microstructure model. Furthermore, macroscopic stress-strain data with different strain amplitudes for calibration of hardening models is made available. A step-by-step guide and detailed documentation is provided. Other researchers are encouraged to set up and test their models with this validation framework.

\section{Conclusions \& Outlook}
The developed validation methodology facilitates fast semi-automatized benchmarks of micromechanical simulations, which is a requirement considering the vast material landscape and the ongoing digitization of materials. Since it relies on deep learning semantic segmentation and large corrected electron backscatter diffraction scans, it allows for statistical benchmarking of actual plasticity traces and crack formation spots. This form of validation can be especially suitable to validate fatigue indicator parameters, while high resolution digital image correlation-based techniques are essential to validate the underlying mechanical fields. The detection of dominant slip traces from micrographs, introduced in \cite{Thomas2020}, can enable automated validation of modeled slip phenomena. Multimodal registration allows for a combination of damage evolution, microtexture, exact damage localization, and prior fatigue specimen reference in one data set while maintaining the environmental testing conditions and VHCF-regime testing capabilities.

The FIP localization is shown not to agree well with the damage locations from the experiments and tends to be overpredictive. This indicates that commonly used FIP are not able to predict the location of fatigue crack initiation reliably, although they might still be applicable for fatigue lifetime prediction.

The acquired data and models are available open source to build the foundation of a more general micromechanical modeling approach. To achieve this, the data set should be extended in the future by more materials, e.g. martensitic steel, and more experimental data such as the 3D microtexture for grain boundary inclination angles or electron channeling contrast imaging for initial statistically-stored dislocation densities.

\section*{Data availability}
The research data is made available in Fordatis \cite{fordatis_repo}. The code for running the simulations and performing the validations is available in a Github repository \cite{github_repo}.




\section*{Author contributions}

\textbf{Ali Riza Durmaz:} Conceptualization, Methodology Development, Software	Programming, Validation	Verification, Formal analysis, Investigation, Data Curation, Writing - Original Draft, Writing - Review \& Editing, Visualization, and Project administration \textbf{Erik Natkowski:} Methodology Development, Software Programming, Validation Verification, Formal analysis, Investigation, Data Curation, Writing - Original Draft, Writing - Review \& Editing, and Visualization \textbf{Nikolai Arnaudov:} Conceptualization, Methodology Development, Software Programming, Validation, Verification, Formal analysis, Investigation, Writing - Original Draft, and Visualization \textbf{Stefan Weihe:} Writing - Original Draft, Writing - Review \& Editing, and Supervision \textbf{Petra Sonnweber-Ribic:} Resources, Writing - Original Draft, Writing - Review \& Editing, and Supervision \textbf{Sebastian Münstermann:} Writing - Original Draft, Writing - Review \& Editing, and Supervision \textbf{Chris Eberl:} Conceptualization, Resources, Writing - Original Draft, Writing - Review \& Editing, and Supervision \textbf{Peter Gumbsch:} Resources, Writing - Original Draft, Writing - Review \& Editing, Supervision, and Funding acquisition.

\section*{Acknowledgements}

The contribution of A.D. was funded by the Bosch-Forschungsstiftung im Stifterverband grant number T113/30074/17.

\nomenclature[0]{$\overline{\left(\cdot\right)}$}{Normalized value}
\nomenclature[1A]{$\khone,\khtwo,\khthree$}{Kinematic hardening parameter}
\nomenclature[1C]{$\CII,\CIZ,\CAA$}{Elasticity parameters of cubic stiffness tensor}
\nomenclature[1m]{$\srs$}{Strain rate sensitivity exponent}
\nomenclature[1M]{$\schmidmat$}{Schmid matrix}
\nomenclature[1Nslip]{$\nslip$}{Number of slip systems}
\nomenclature[1n]{$\slipnormal$}{Slip system normal}
\nomenclature[1m_zz_]{$\slipdirection$}{Slip system direction}
\nomenclature[1FIPP]{$\fipp$}{Accumulated plastic slip}
\nomenclature[1FIPFS]{$\fipfs$}{Fatemi-Socie FIP}
\nomenclature[1FIPW]{$\fipw$}{Dissipated energy FIP}
\nomenclature[1FIPINT]{$\fipint$}{Intergranular FIP}
\nomenclature[1t]{$\tme$}{Time}
\nomenclature[1k]{$\kfs$}{Fatemi-Socie FIP constant}
\nomenclature[1kint]{$\kint$}{Intergranular FIP constant}
\nomenclature[1u]{$\dispthree$}{Macroscopic out-of-plane displacement}
\nomenclature[1c]{$\surfintervec$}{Surface intersection vector}
\nomenclature[1nsurf]{$\normalsurf$}{Surface normal vector}
\nomenclature[1deq]{$\deq$}{Equivalent grain diameter}
\nomenclature[1R]{$\rotmat$}{Roatation matrix}
\nomenclature[1f0]{$f_0$}{Resonant frequency}
%
\nomenclature[2a]{$\slipsysind$}{Slip system index}
\nomenclature[2amax]{$\slipsysindmax$}{Slip system index for maximum FIP value}
\nomenclature[2e]{$\strain$}{Small strain tensor}
\nomenclature[2ep]{$\strainplastic$}{Plastic strain}
\nomenclature[2c]{$\gamma$}{Shear}
\nomenclature[2cnet]{$\shearnet$}{Net shear}
\nomenclature[2cref]{$\shearrateref$}{Reference shear rate}
\nomenclature[2cp]{$\plasticshear$}{Plastic shear}
\nomenclature[2D]{$\Dshearmax$}{Plastic shear range}
\nomenclature[2Df]{$\Delta f$}{Frequency change}
\nomenclature[2v]{$\backstress$}{Backstress}
\nomenclature[2s]{$\rss$}{Resolved shear stress}
\nomenclature[2sc]{$\crss$}{Critical resolved shear stress}
\nomenclature[2rmax]{$\normalstressmax$}{Maximum normal stress}
\nomenclature[2rgb]{$\stressgb$}{Average peak normal stress on grain boundary}
%
\nomenclature[3CP]{CPFE}{Crystal plasticity finite element method }
\nomenclature[3EBSD]{EBSD}{Electron backscatter diffraction }
\nomenclature[3FIP]{FIP}{Fatigue indicator parameter }
\nomenclature[3SEM]{SEM}{Scanning electron microscopy}
\nomenclature[3HRDIC]{HR-DIC}{High-resolution digital image correlation}
\nomenclature[3CNN]{CNN}{Convolutional neural network}
\nomenclature[3WD]{WD}{Working distance}
\nomenclature[3SE2]{SE2}{Secondary electron}
\nomenclature[3LCF]{LCF}{Low cycle fatigue}
\nomenclature[3VHCF]{VHCF}{Very high cycle fatigue}
\nomenclature[3BCC]{BCC}{Body centered cubic}
\nomenclature[3GB]{GB}{Grain boundary}
\nomenclature[3ROC]{ROC}{Receiver operator characteristic}
\nomenclature[3AUC]{AUC}{Area under curve}
\printnomenclature


\newpage
\bibliography{bib_combined}

\end{document}